\documentclass[useAMS,usegraphicx,usenatbib]{mn2e}
\usepackage{url}

\title[Cosmic sulfur species]{Sulfur-bearing species in molecular clouds}
\author[G. Bilalbegovi\' c, G. Baranovi\' c]
{G. Bilalbegovi\' c$^{1}$,
G. Baranovi\' c$^{2}$
\\
$^{1}$Department of Physics, Faculty of Science, University of Zagreb, Bijeni\v cka 32, 10000 Zagreb, Croatia\\
$^{2}$Rudjer Bo\v skovi\' c Institute, Division of Organic Chemistry and Biochemistry,
Bijeni\v  cka 54, 10000 Zagreb, Croatia
}

\begin{document}

\date{\today}

\pagerange{\pageref{firstpage}--\pageref{lastpage}}
\pubyear{2013} \volume{000}

\maketitle 
\label{firstpage}

\begin{abstract}
We study  several molecules that
could help in the solution of the missing sulfur problem in dense clouds and  circumstellar regions, as well as in the clarification of the sulfur chemistry in comets.
These sulfur molecules are: the trimer (CH$_2$S)$_3$ and the tetramer (CH$_2$S)$_4$ of thioformaldehyde, pentathian S$_5$CH$_2$, hexathiepan S$_6$CH$_2$, thiirane C$_2$H$_4$S, trisulfane HSSSH, and thioacetone (CH$_3$)$_2$CS.
Infrared spectra of these species are calculated using density functional theory methods. The majority of calculated bands belong to the mid-infrared, with some of them occurring  in the near and far-infrared region.
We suggest that some of unidentified  spectral features measured by {\it Infrared Space Observatory}
in several active galactic nuclei and starburst  galaxies
could be caused by 1,3,5-trithiane ((CH$_2$S)$_3$), 1,3,5,7-tetrathiocane ((CH$_2$S)$_4$), and thiirane (C$_2$H$_4$S).
The objects whose unidentified infrared features we compare with calculated bands are: NGC 253, M82, NGC 1068, Circinus, Arp 220, 30 Doradus,
Orion KL, and Sgr B2.

\end{abstract}


\section{Introduction}
\label{intro}

More than 180 molecules  have been discovered in space \citep{Tielens2013}.
With an exception of fullerenes, the largest molecule confirmed to exist
in space till now consist of thirteen atoms \citep{Bell1997}.
Several  sulfur-bearing molecules and ions have been observed in the interstellar medium and circumstellar shells in the Milky Way, as well as
in other galaxies. 
The ethyl mercaptan, CH$_3$CH$_2$SH, the largest astrophysical 
molecule containing sulfur, was recently observed in Orion \citep{Kolesnikova2014}.
Interstellar methyl mercaptan, CH$_3$SH, 
was discovered long time ago by measuring lines in the millimeter range of the spectrum toward the Sgr B2 \citep{Linke1979}.
Methyl mercaptan was also detected in other environments, for example in the organic-rich hot core G327.3-0.6 \citep{Gibb2000}, and in Orion
\citep{Kolesnikova2014}.
It was suggested that methyl mercaptan forms in interstellar ices and then evaporates in hot cores \citep{Gibb2000}.

Although the first interstellar sulfur-bearing molecule CS was detected a long time ago \citep{Penzias1971},  
the chemistry of interstellar sulfur is still undetermined.
The observed abundance of sulfur-bearing species in dense clouds
is only about 0.1$\%$  of the same quantity in diffuse clouds 
\citep{Tieftrunk1994,Charnley1997}. 
Therefore, the main sulfur species  in dense regions of
interstellar medium are still unknown.  Molecules  OCS  \citep{Palumbo1997}
and SO$_2$ \citep{Zasowski2009} were detected in interstellar ices, but their abundances are not sufficient to account 
for the missing sulfur.
Various species  have been suggested as a possible reservoir
of sulfur, for example  polysulfanes H$_2$S$_n$, sulfur polymers S$_n$, solid H$_2$SO$_4$, and FeS
\citep{Druard2012,Scappini2003,Keller2002}.
The presence of S$^+$ and its efficient sticking to dust grains has also been proposed \citep{Ruffle1999}, as well as the existence of a substantial amount of atomic sulfur in shocked gases \citep{Anderson2013}.

Sulfur species have also been observed in comets \citep{Comets2004}. A cometary ice of the  Hale--Bopp (C/1995 O1) had a
similar IR (infrared) spectrum as interstellar ices in dense clouds \citep{Crovisier1997,Bockelee2000}.
Molecules H$_2$S, CS, OCS, H$_2$CS, and SO were observed in the comet Hale--Bopp using radio telescopes \citep{Woodney1997,Biver1997}.  
The low abundance of H$_2$CS in comparison to H$_2$S, as well as large daily variations in profiles and intensities of lines for H$_2$S and CS were found \citep{Woodney1997}.  Therefore, an intensive sulfur chemistry was in progress when  the comet Hale--Bopp was observed from Earth.

Models of interstellar ices containing H$_2$S, SO$_2$, CO, CO$_2$, CH$_3$OH, H$_2$O, and/or  NH$_3$ were studied in laboratories on Earth by protons irradiation \citep{Ferrante2008,Garozzo2010}, under UV \citep{Jimenez2011,Jimenez2014}, and soft X-rays \citep{Escobar2012}.
The effects of irradiation and of thermal processing were explored by measuring IR and mass spectra. It was found that the amount of 
H$_2$S, SO$_2$, and CO decreased and  new molecules appeared. Depending on the initial ice composition, the radiation dose, and the temperature, some of
detected molecules were: OCS, SO$_3$,  O$_3$, H$_2$S$_2$, HS$_2$, CS$_2$, H$_2$CO, CH$_4$,
CO$_2$, S$_n$, n=2-8. A sulfur rich residuum without IR bands also formed 
when the ice containing H$_2$S was irradiated by protons \citep{Garozzo2010}.

The thioformaldehyde molecule, H$_2$CS, in our galaxy was detected by measuring a higher transition 
2$_{11} \leftarrow$ 2$_{12}$ at 3139.4 MHz, in the direction of Sgr B2 \citep{Sinclair1973}.   H$_2$CS was observed
in cold and hot  cores, outflows of stars,  and in the interstellar medium \citep{Maeda2008}. 
Thioformaldehyde was also observed in other galaxies \citep{Heikkila1999,Martin2005}.
The H$_2$CS molecule is unstable under laboratory conditions on Earth \citep{Peach1975}. It is prone to polymerization and, as possible products, various cyclic molecules might occur. Among simplest ones are (CH$_2$S)$_3$ (1,3,5-trithiane) and  (CH$_2$S)$_4$ (1,3,5,7-tetrathiocane).  
Properties of  higher polymeric thioformaldehydes, (CH$_2$S)$_n$, have also been studied on Earth \citep{Credali1967,Peach1975}.
The cyclic structures, (CH$_2$S)$_n$, are stable.  Thus, they appear as a possible reservoir of sulfur in space. 
In laboratories (CH$_2$S)$_3$ forms when formaldehyde and hydrogen sulfide react.  
Various routes and catalysts for this reaction have been found.
One way to form (CH$_2$S)$_3$ is the reaction between H$_2$CO and H$_2$S under irradiation \citep{Credali1967}.
Both H$_2$CO and H$_2$S have been discovered in our and other galaxies
\citep{Snyder1969,Gardner1974,Thaddeus1972,Heikkila1999}.
It was proposed that formaldehyde forms in interstellar ices by hydrogenation of CO \citep{Tielens1997,Watanabe2002}. 
The observed abundances of H$_2$S are such that it has been assumed that this molecule also forms in interstellar ices on
dust grains \citep{Charnley1997,Buckle2003}.
We expect that, as in chemical laboratories on Earth, 
formaldehyde and hydrogen sulfide react in space under irradiation (such as UV, X, or cosmic rays), 
and that (CH$_2$S)$_3$ and higher polymeric thioformaldehydes are present. 
Formaldehyde and hydrogen sulfide \citep{Biver1997}, as well as thioformaldehyde \citep{Woodney1997},
were also detected in the Hale--Bopp.   In addition,  several other sulfur-bearing species and changes of intensities for H$_2$S  during observations
were detected in this comet.  It is possible to suggest that formaldehyde and hydrogen sulfide react in the Hale-Bopp (and perhaps in some other comets) and form cyclic thioformaldehydes.

Cyclic molecules pentathian (S$_5$CH$_2$) and hexathiepan (S$_6$CH$_2$) were detected in laboratory astrophysics experiments on interstellar ices after the UV irradiation at T= 12 K, which was followed by  heating to room temperature
\citep{Guillermo2002}. Mass spectra were measured and, among pentathian, hexathiepan and some known molecules, 
several unidentified species were found.
This opens a possibility that (CH$_2$S)$_3$, (CH$_2$S)$_4$ and other sulfur-bearing species were also present, or could formed under suitable conditions.
Disulfane, H$_2$S$_2$, was also detected in laboratory experiments on interstellar ices 
\citep{Ferrante2008,Garozzo2010,Jimenez2011,Escobar2012,Jimenez2014}. The presence of polysulfanes, H$_2$S$_n$,  was suggested \citep{Druard2012}. The results of calculations on chemical networks  predicted that
H$_2$S$_3$ is more abundant than H$_2$S$_2$ in less dense clouds at T= 20 K
\citep{Druard2012}.  Therefore, H$_2$S$_3$ is an important molecule for further studies of sulfur species in molecular clouds.
We investigate the most stable conformation {\it trans}-HSSSH \citep{Liedtke1993}.
Thioacetone, (CH$_3$)$_2$CS, in the form we study, consists of the two methyl groups  and the sulfur atom  doubly bonded to carbon that could give rise to specific IR spectral features. Thioacetone could exist as the thioketo and thienol tautomers, but thioenol transforms to the thioketo form \citep{Lipscomb1970}.
We study the thioketo tautomer that is prone to polymerization,
and therefore unstable. However, its lifetime is several minutes \citep{Kroto1974}. 
Thiirane (C$_2$H$_4$S, ethylene sulfide) is the smallest cyclic molecule containing sulfur. Its oxygen analogue oxirane (C$_2$H$_4$O) is
already confirmed as an astrophysical molecule \citep{Dickens1997,Puzzarini2014}.

Most sulfur-bearing and other astrophysical molecules have been detected by radio telescopes involving rotational spectra.   
For some of molecules studied here by calculating their infrared spectra, rotational spectra and constants are available: 
(CH$_2$S)$_3$ \citep{Antolinez2002, Cervellati1984, Dorofeeva1995},  trans-HSSSH \citep{Liedtke1993,Liedtke1997}, (CH$_3$)$_2$CS \citep{Kroto1974},  C$_2$H$_4$S \citep{Hirao2001,Kirchner1997,Hirose1976}.

In this work, to shed light on  the missing  reservoirs of sulfur in dense interstellar clouds, 
we study several sulfur-bearing species. This study is also important for understanding of the sulfur chemistry in comets. IR spectra of sulfur molecules are calculated using density functional theory methods \citep{Becke2014}.
In Sec. 2 we describe computational methods. In Sec. 3.1 we present the calculated properties of molecules and their IR spectra. We compare calculated bands with observed spectra of astrophysical objects and find good agreement between some calculated bands and unidentified features observed by {\it Infrared Space Observatory (ISO)} in several starburst and active galaxies \citep{UnidentifiedISO,Sturm2000,Lutz2000,Fischer1999}. Therefore, in Sec. 3.2 we present the list of already detected sulfur-bearing species in molecular gases of these galaxies and briefly describe conditions for their formation.  In Sec. 3.3 we list detected sulfur-bearing species in environments of our galaxy: Orion KL and Sgr B2.   
The conclusions are summarized in Sec. 4.

\section{Computational methods}
\label{methods}

We use density functional theory methods within the Gaussian code \citep{Gauss09} to study several sulfur-bearing molecules.
The hybrid B3LYP functional \citep{Becke1993,Stephens1994} and the augmented correlation consistent 
aug-cc-pVTZ  basis set  
\citep{Woon1993}  are applied. It was found that this basis set and the B3LYP functional produced accurate IR intensities
\citep{Zvereva2011}.
Using these methods we compute harmonic vibrational frequencies.

The Gaussian code and B3LYP functional have  been successfully applied for studies of IR spectra of interstellar molecules and nanoparticles
\citep{Bauschlicher2010,Boersma2014,Ricca2012,Goumans2011}.
Scale factors are used in this type of calculations \citep{Bauschlicher2010} to minimize errors due to anharmonic effects, basis set incompleteness and unknown exchange-correlation effects in density functional theory approximations. 
The other, more rigorous, way is to do vibrational calculations with anharmonic correction to frequencies and intensities. 
We tested a new implementation of full anharmonic IR intensities at the second-order of perturbation theory level 
\citep{Bloino2012,Barone2010}. However, we found that the present implementation of this method (in the Gaussian 09 code)
is not suitable for the C$_{3v}$ symmetry of (H$_2$CS)$_3$.

In this work
nine molecules (CS$_2$ \citep{Shimanouchi1972}, C$_2$H$_4$S  \citep{Allen1986}, C$_2$H$_4$O  \citep{Puzzarini2014}, HS$_2$H \citep{Jacox1994}, H$_2$CS \citep{Jacox1994}, CH$_3$SH \citep{Sverdlov1974}, CH$_3$SCH$_3$ \citep{Sverdlov1974}, SO$_2$ \citep{Person1982}, and H$_2$S \citep{Shimanouchi1972}) were used to obtain the value of 0.9687 for the scaling factor. The mean average error, the mean absolute error, and the root mean square error were 7.7, 18.9 and 23.1 cm$^{-1}$, respectively. 
The calculated intensities for the nine molecules were in good agreement with experiment what was actually expected from the chosen level of theory (B3LYP/aug-cc-pVTZ). 
The calculated frequencies of (CH$_2$S)$_3$, (CH$_2$S)$_4$, S$_5$CH$_2$, S$_6$CH$_2$, HSSSH, (CH$_3$)$_2$CS, and C$_2$H$_4$S were subsequently scaled with 0.9687
and, together with calculated intensities, used for astrophysical
predictive purposes. For IR spectra presented in this work we used  Lorentzian band profiles with FWHM of 15 cm$^{-1}$.

\section{Results and Discussion}
\label{results}

\subsection{Infrared Spectra}
\label{ir}

\begin{figure}
\centering 
\includegraphics[scale=0.2]{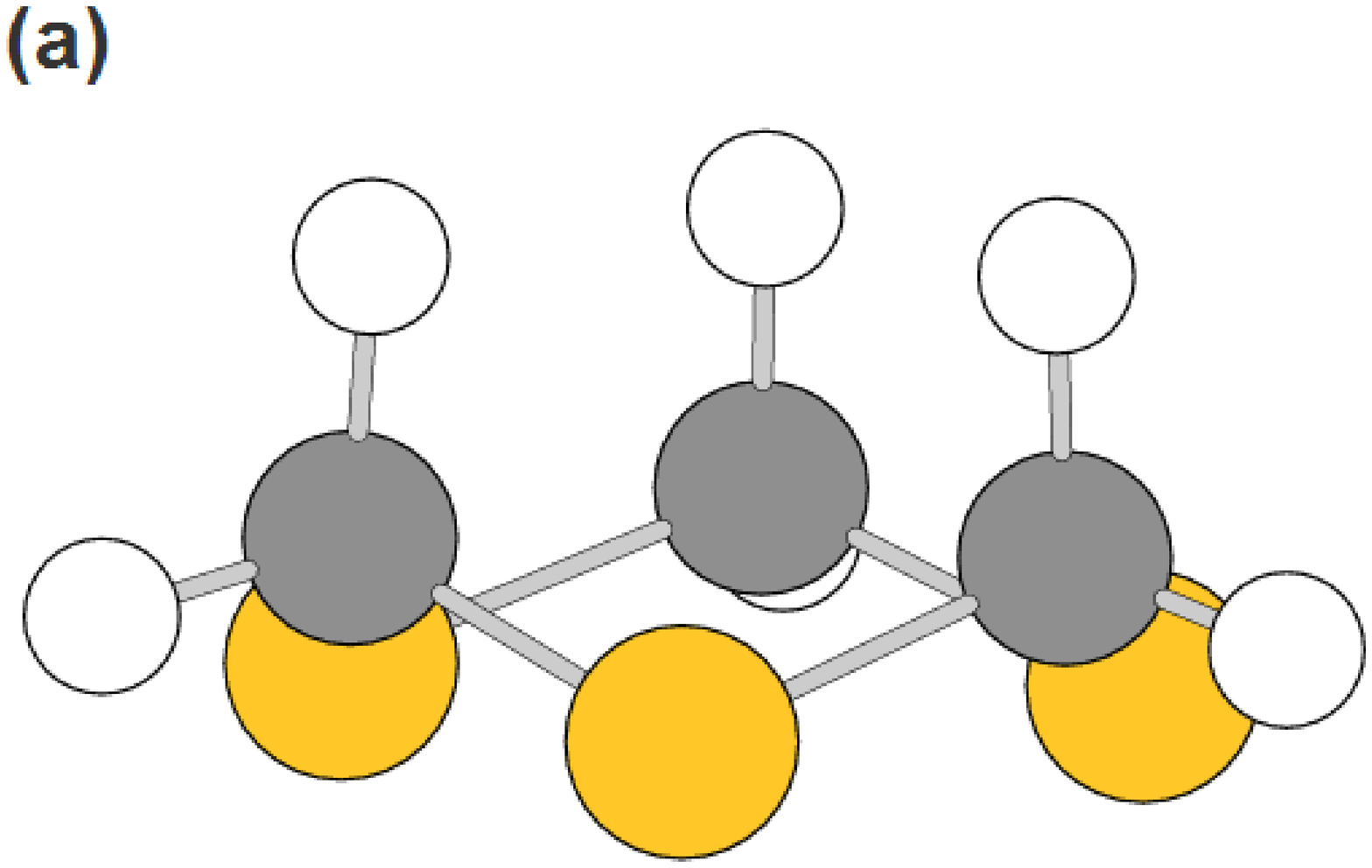}

\includegraphics[scale=0.2]{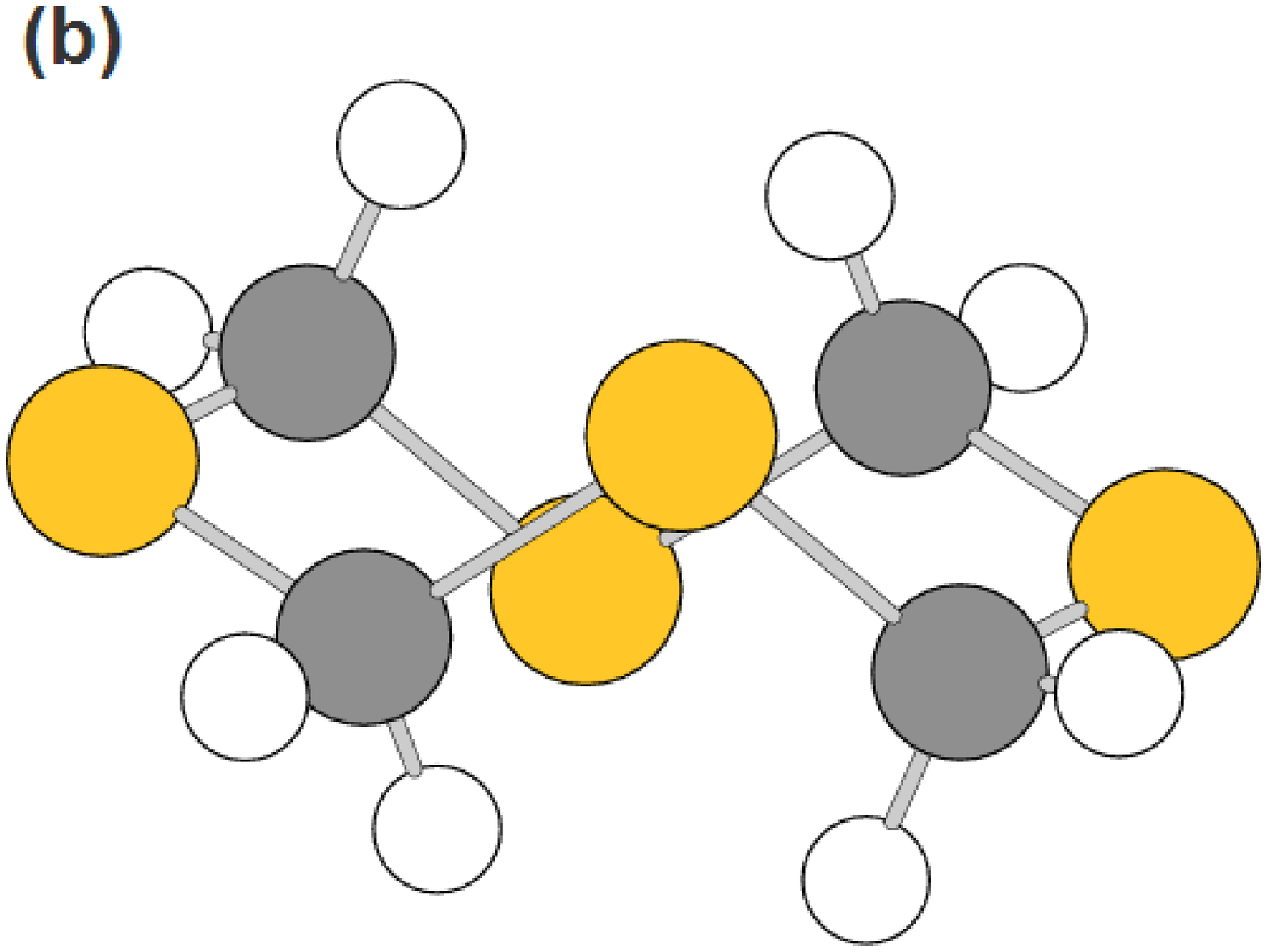}
\caption{
The optimized structures: (a) (CH$_2$S)$_3$ (thioformaldehyde trimer, 1,3,5-trithiane), 
(b) (CH$_2$S)$_4$ (thioformaldehyde tetramer, 1,3,5,7-tetrathiocane).
Small white balls represent hydrogen atoms, light gray (yellow in the color figure) are for sulfur atoms, and dark gray for carbon atoms. }
\label{fig1}
\end{figure}

\begin{figure}
\centering 
\includegraphics[scale=0.2]{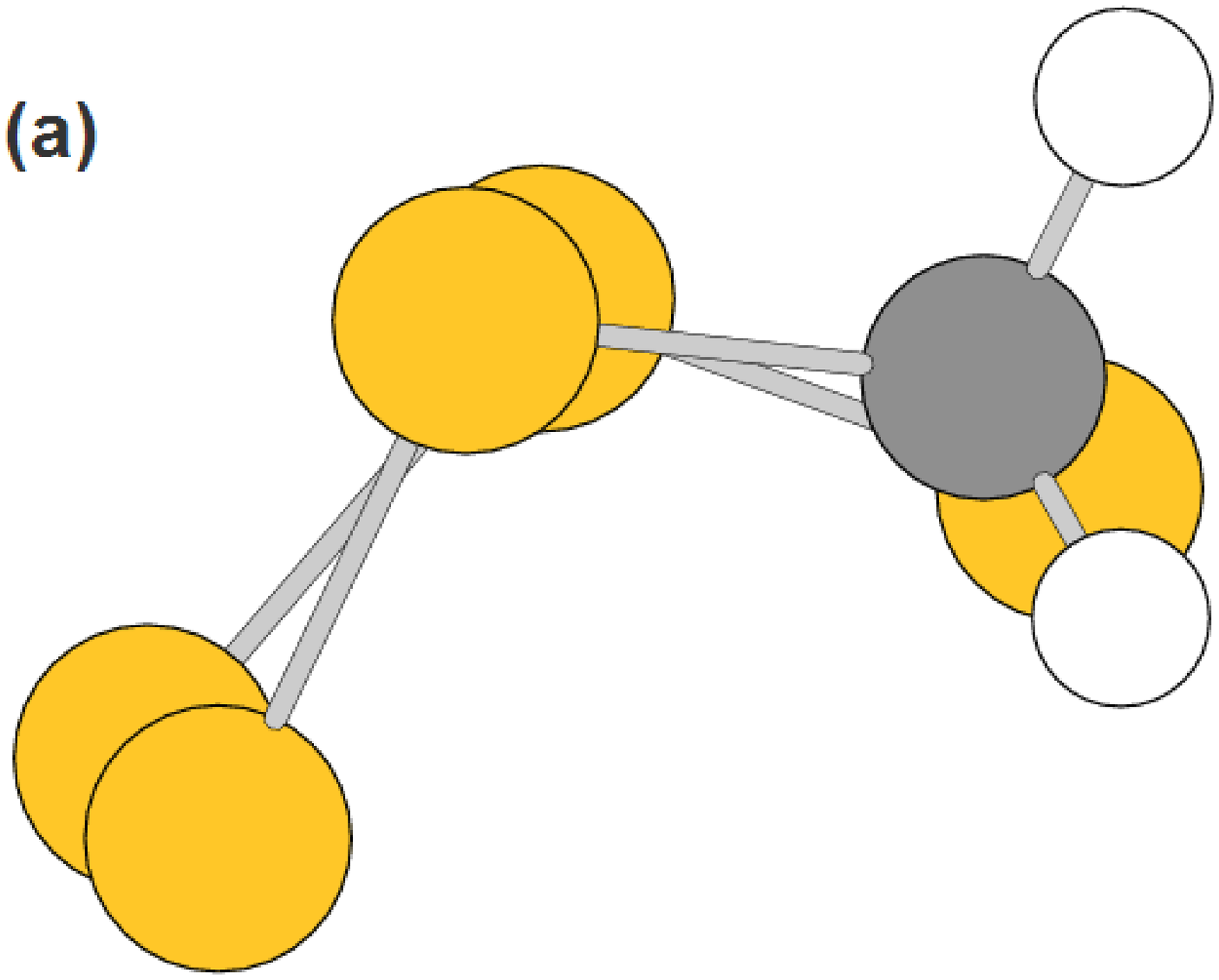}

\includegraphics[scale=0.2]{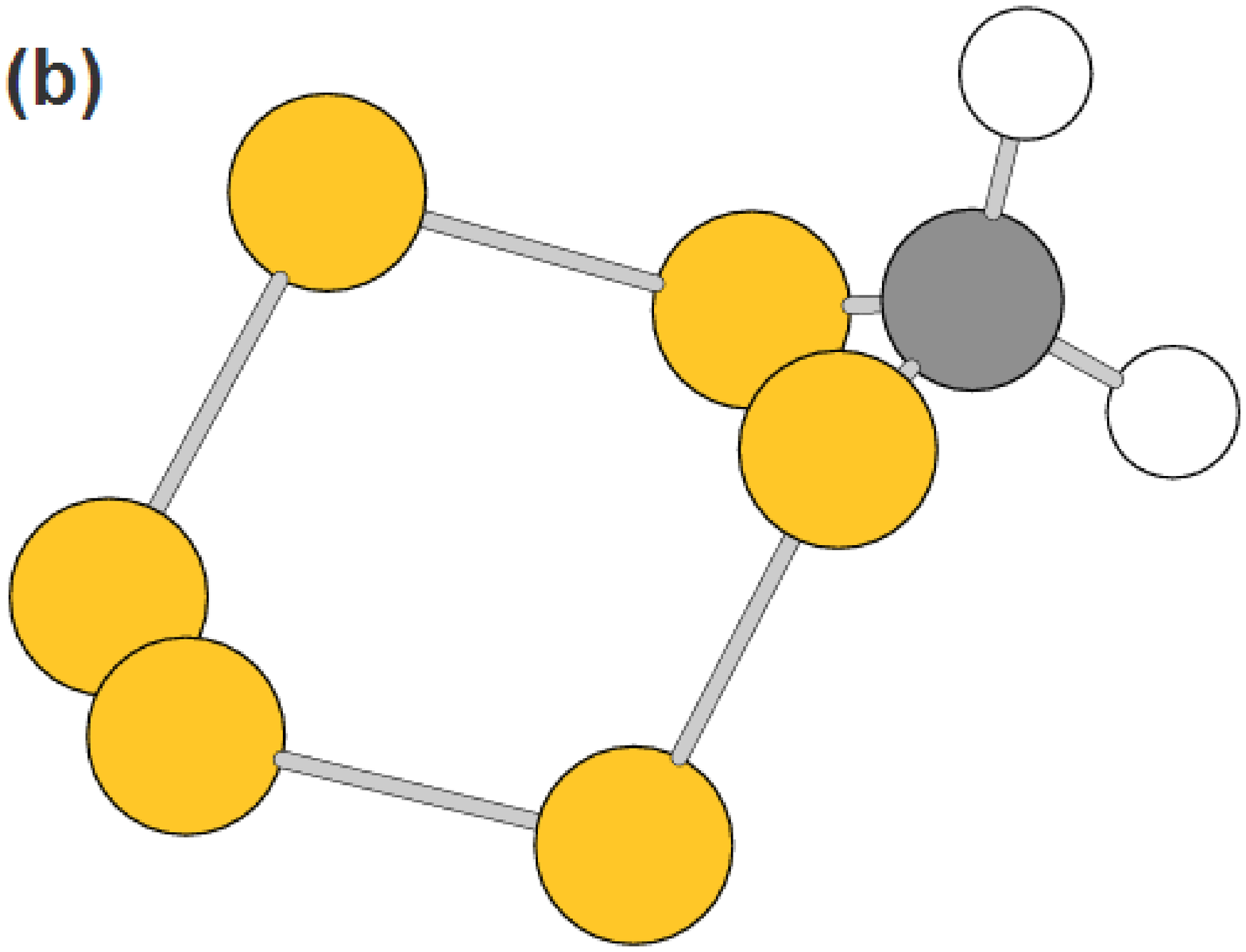}
\caption{
The optimized structures: (a) S$_5$CH$_2$ (pentathian), 
(b) S$_6$CH$_2$ (hexathiepan).
Details as in Fig.~\ref{fig1}.}
\label{fig2}
\end{figure}

\begin{figure}
\centering 
\includegraphics[scale=0.2]{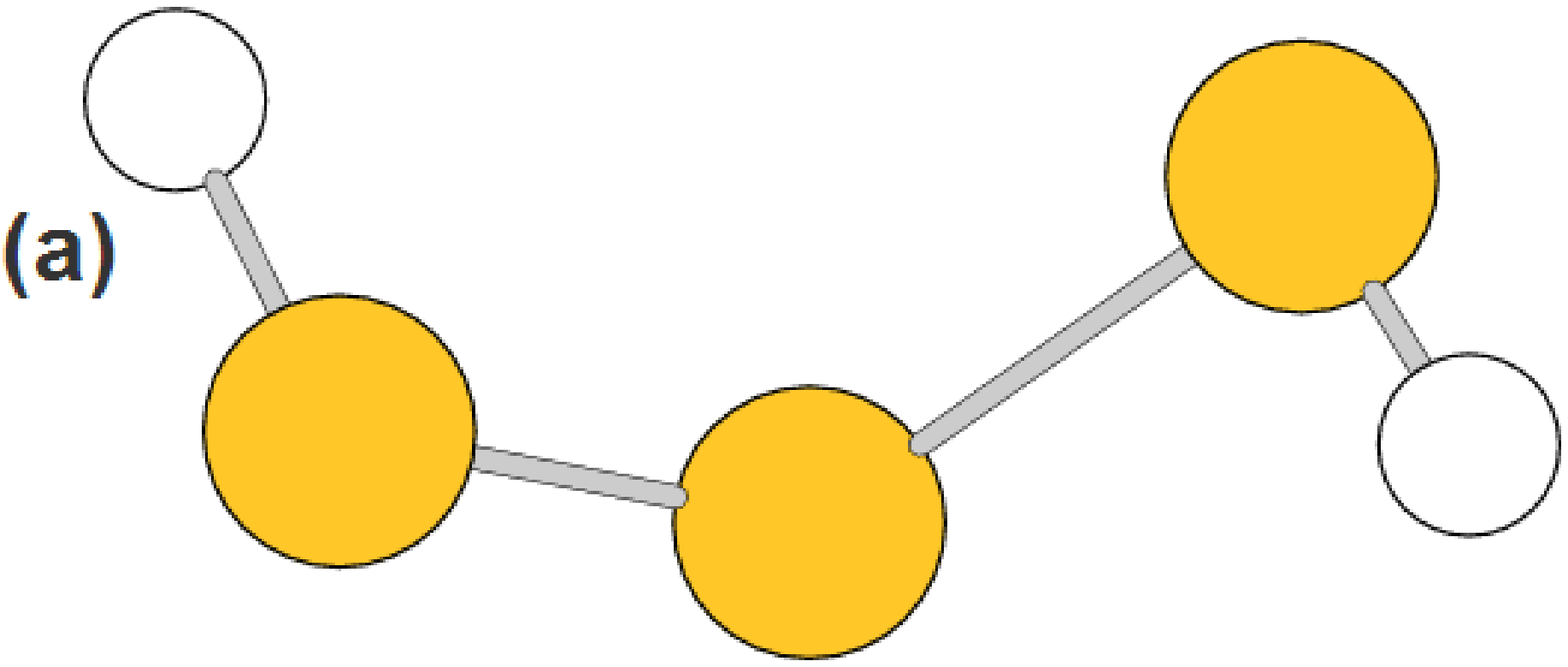}

\includegraphics[scale=0.2]{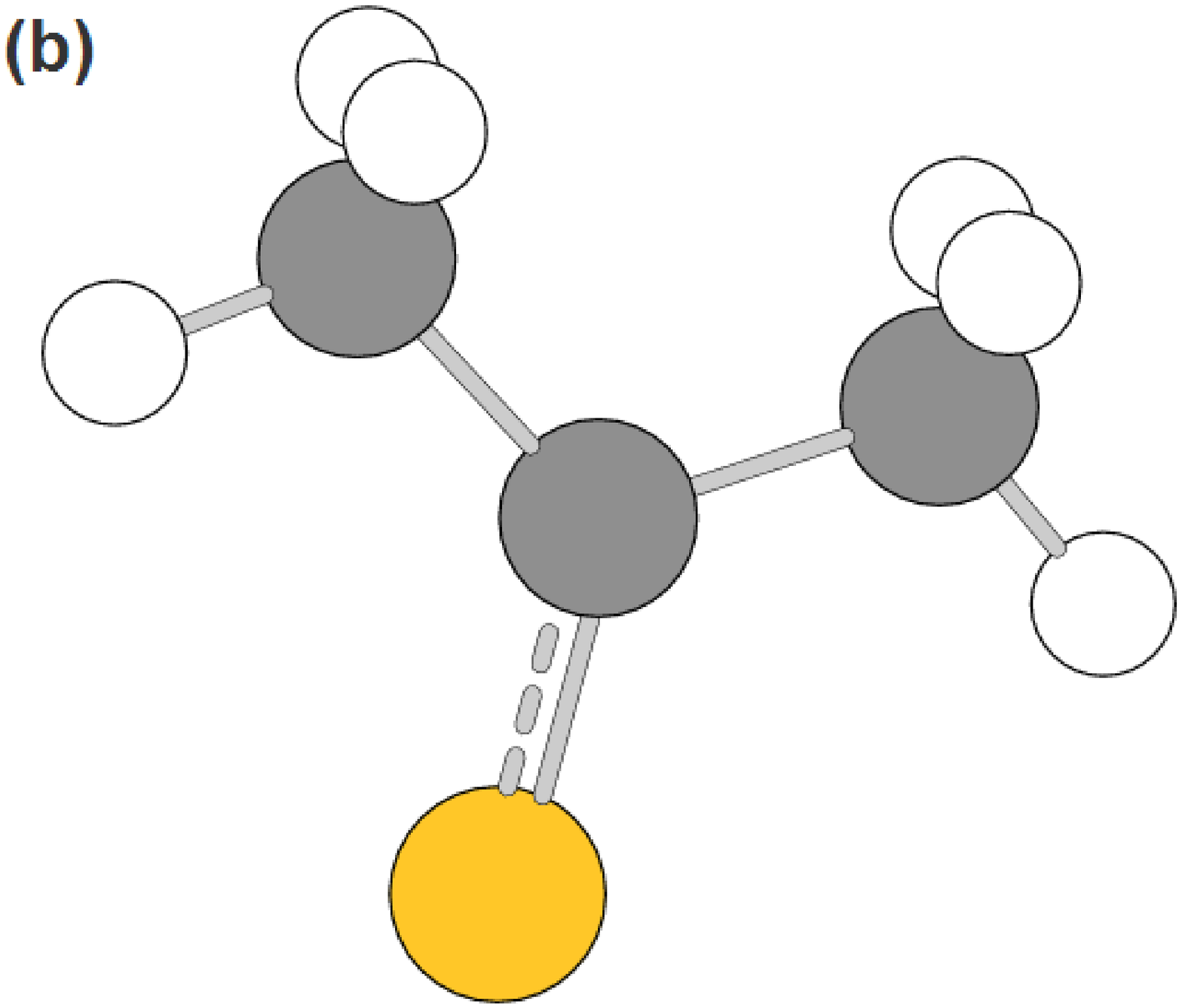}

\includegraphics[scale=0.2]{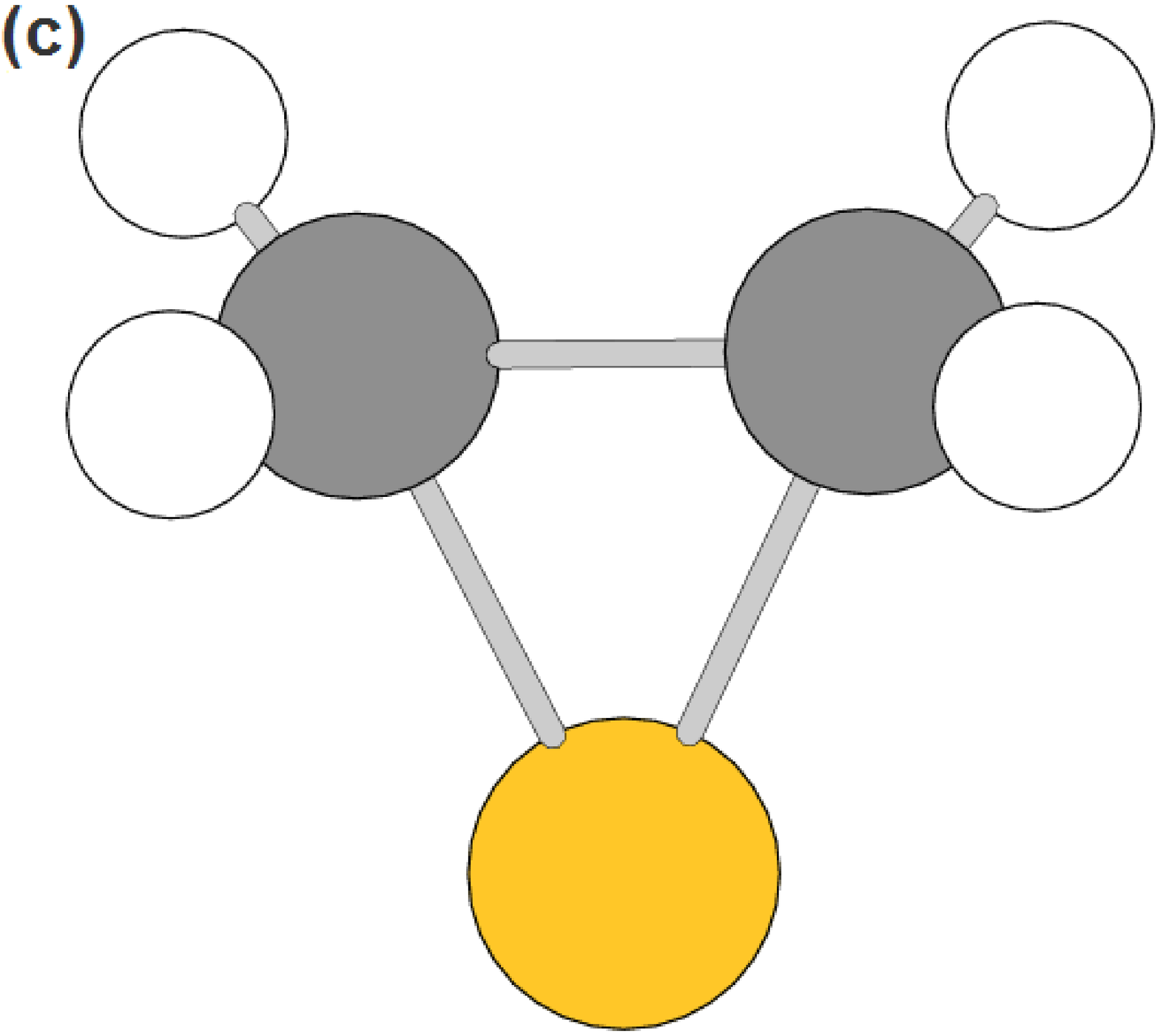}
\caption{
The optimized structures: (a) {\it trans}-HSSSH, 
(b) (CH$_3$)$_2$CS (thioacetone),
(c) C$_2$H$_4$S (thiirane). 
A double C-S bond in (CH$_3$)$_2$CS is shown.
Details as in Fig.~\ref{fig1}.}
\label{fig3}
\end{figure}

Figs. 1-3 show optimized geometries of molecules we study.
Both (CH$_2$S)$_3$ and (CH$_2$S)$_4$  are chair-shaped cyclic structures (Fig.~\ref{fig1}), where rings are composed of, respectively,  six and eight alternating C and S atoms.
In contrast, in the rings of S$_5$CH$_2$ and  S$_6$CH$_2$ (Fig.~\ref{fig2}) only one carbon atom exists, whereas two C atoms appear in 
the cyclic structure of C$_2$H$_4$S (Fig.~\ref{fig3} (c)). Zig-zag (or helix) structures of either S, or C atoms exist for, respectively,
HSSSH (Fig.~\ref{fig3} (a)) and  C$_3$H$_6$S (Fig.~\ref{fig3} (b)).
The coordinates of optimized structures for all molecules are available as the online-only supporting information.

We have calculated that (CH$_2$S)$_3$ has a large dipole moment of 2.1787  Debye.
In contrast the dipole moment of (CH$_2$S)$_4$ is zero. It is known that molecules with a zero dipole moment are radio inactive. Therefore, (CH$_2$S)$_4$ 
could only be detected in the IR spectral region and
for radio telescopes this molecule would  behave as the hidden part of a possible reservoir of sulfur.
Dipole moments of 1.9784, 1.7086, 1.9669, 0.5085, 2.9493 Debye are calculated
for S$_5$CH$_2$, S$_6$CH$_2$, C$_2$H$_4$S, HSSSH, (CH$_3$)$_2$CS, respectively.

\begin{figure}
\centering 
\includegraphics[scale=0.4]{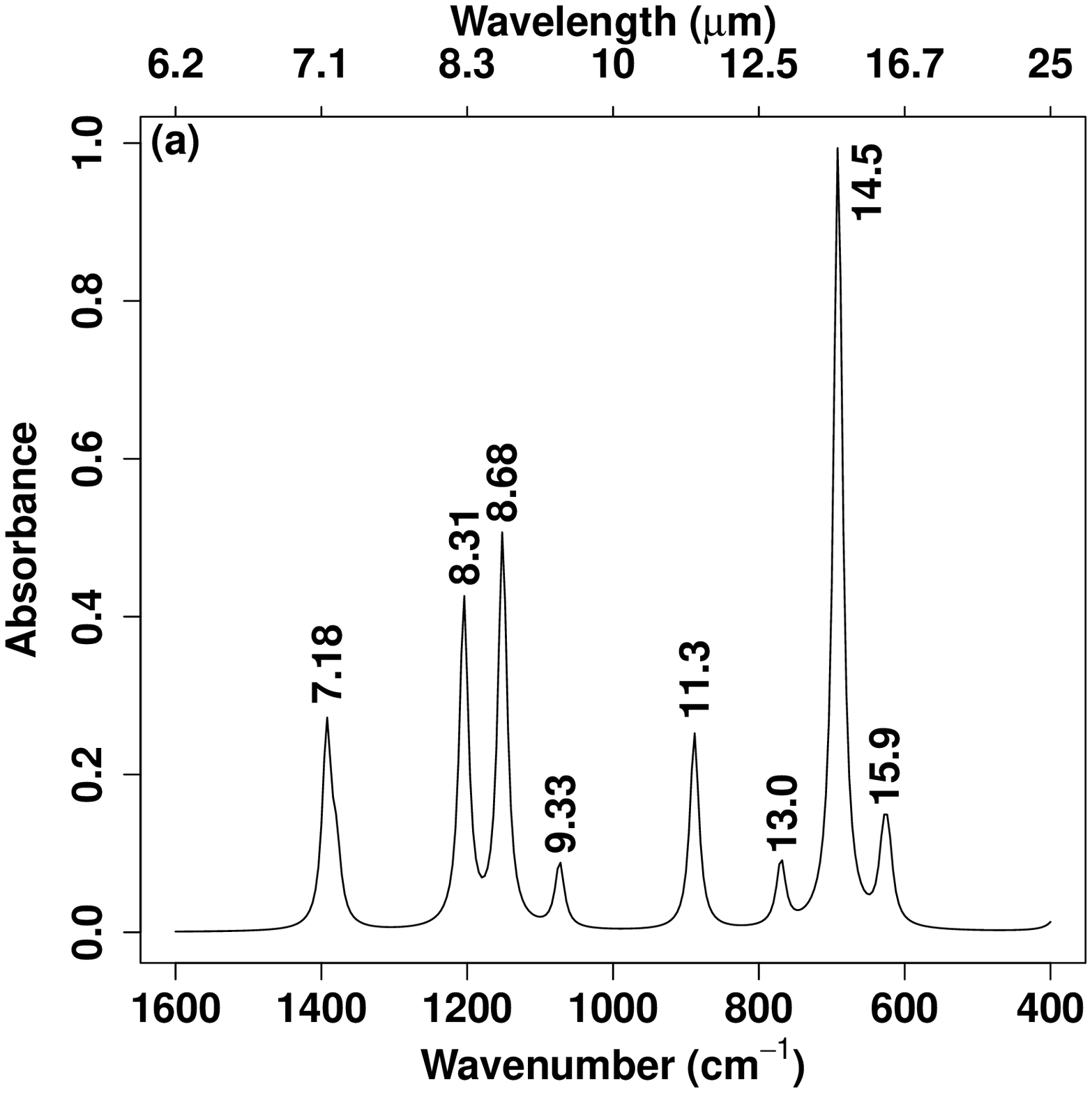}

\includegraphics[scale=0.4]{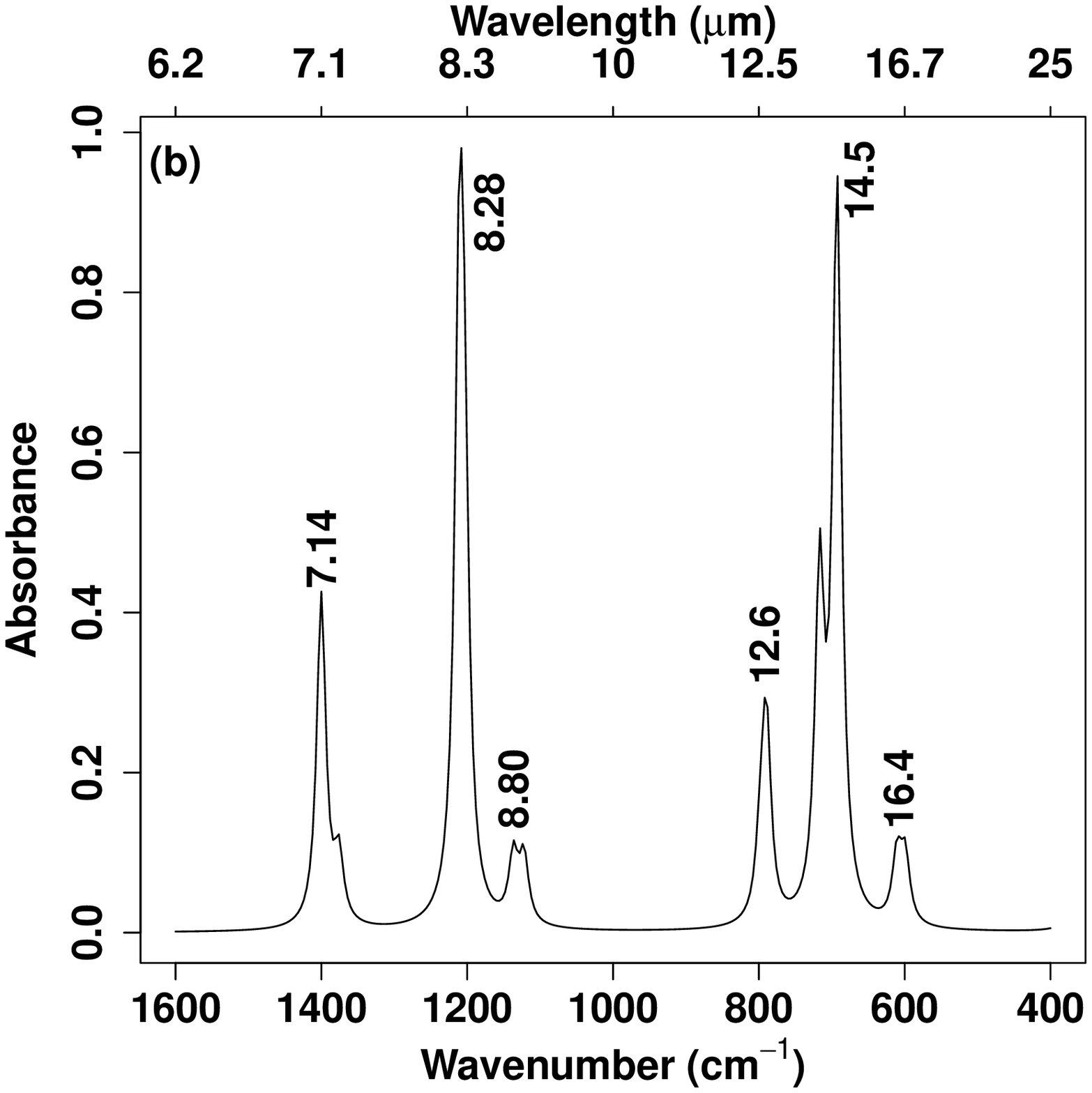}
\caption{IR spectrum: (a) (CH$_2$S)$_3$, (b) (CH$_2$S)$_4$. Positions at the peaks are in $\mu$m.}
\label{fig4}
\end{figure}

\begin{figure}
\centering 
\includegraphics[scale=0.4]{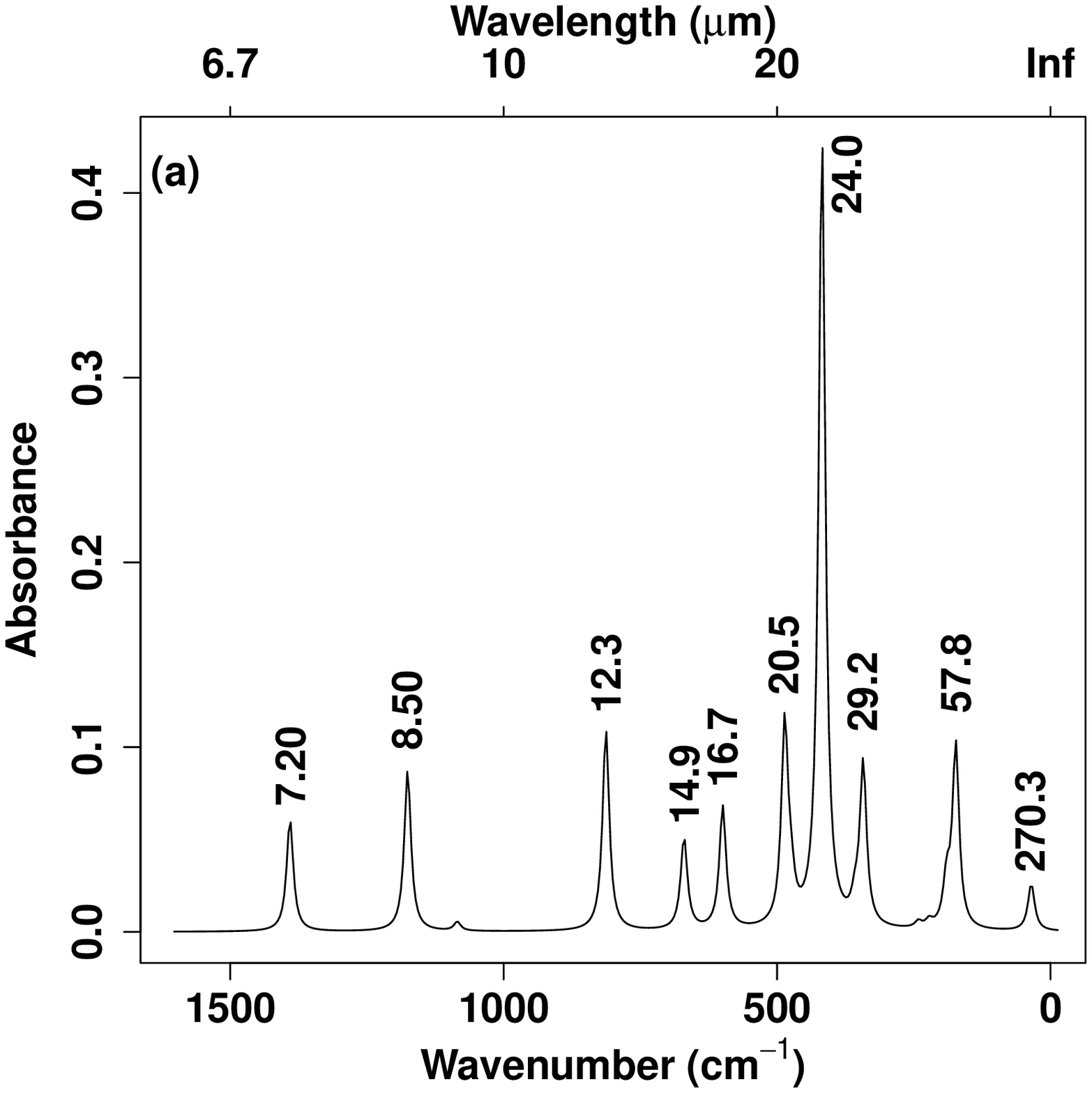}

\includegraphics[scale=0.4]{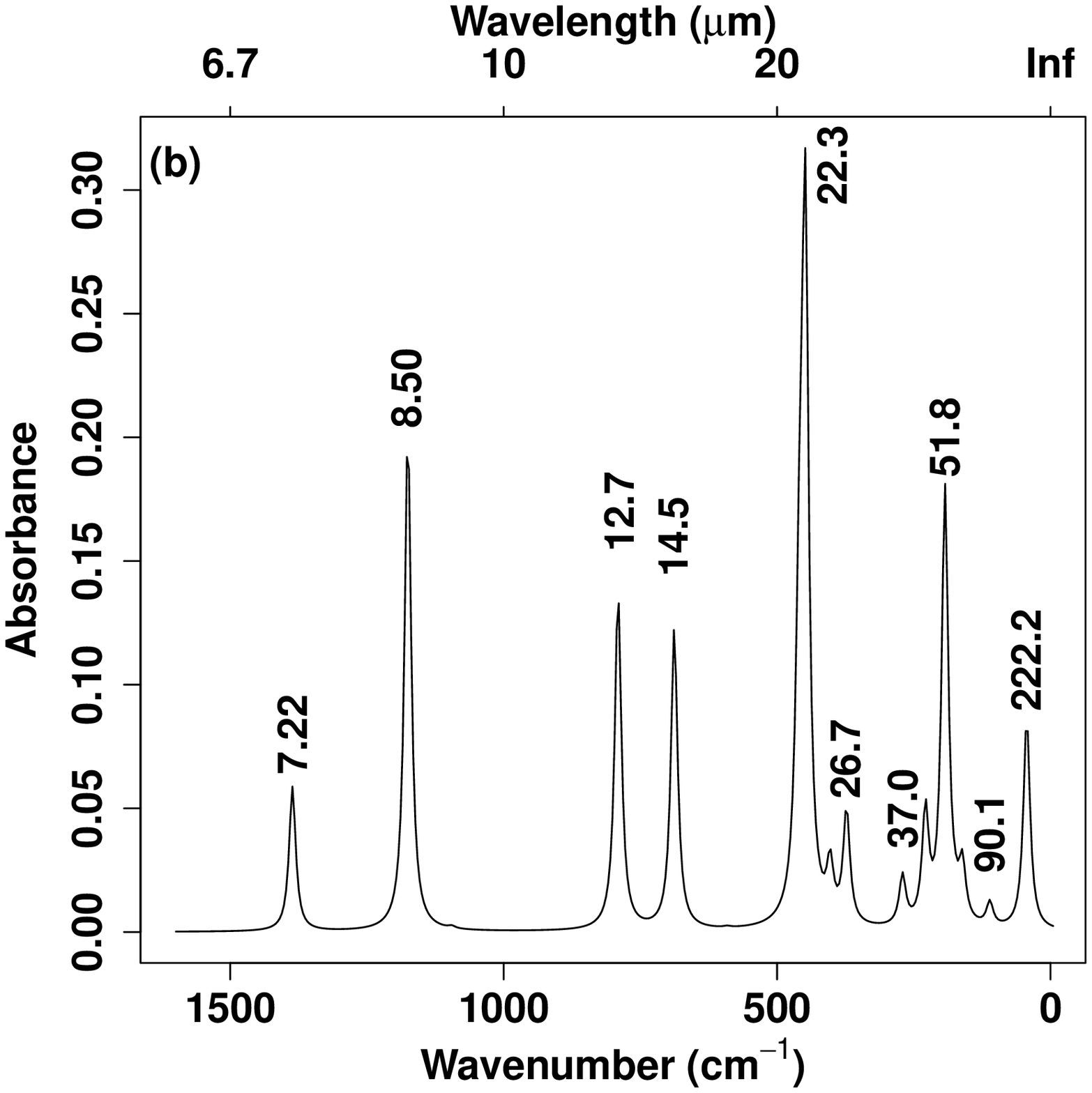}
\caption{IR spectrum  of:  (a) S$_5$CH$_2$, 
(b) S$_6$CH$_2$. Positions at the peaks are in $\mu$m.}
\label{fig5}
\end{figure}

\begin{figure}
\centering 
\includegraphics[scale=0.4]{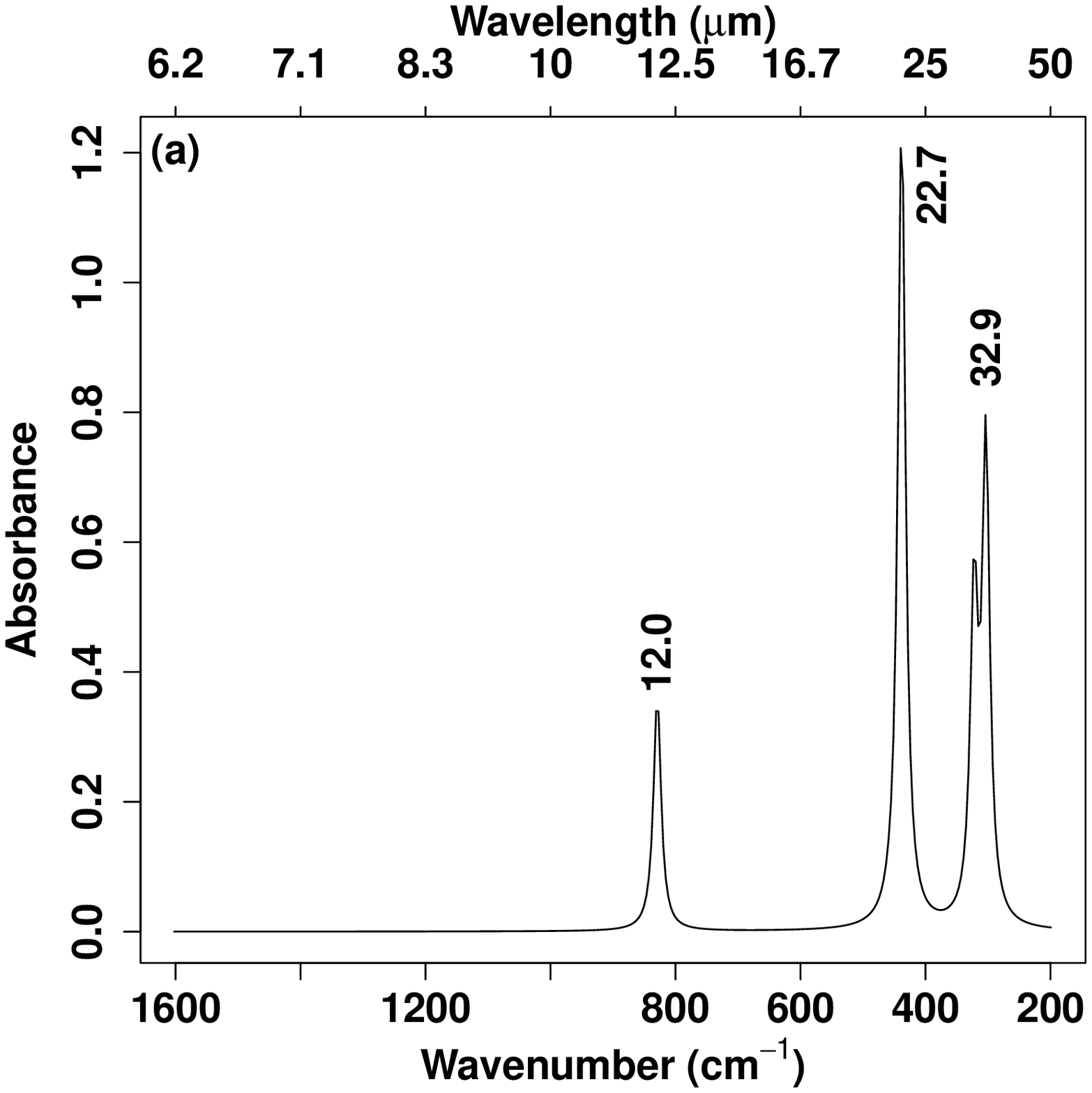}

\includegraphics[scale=0.4]{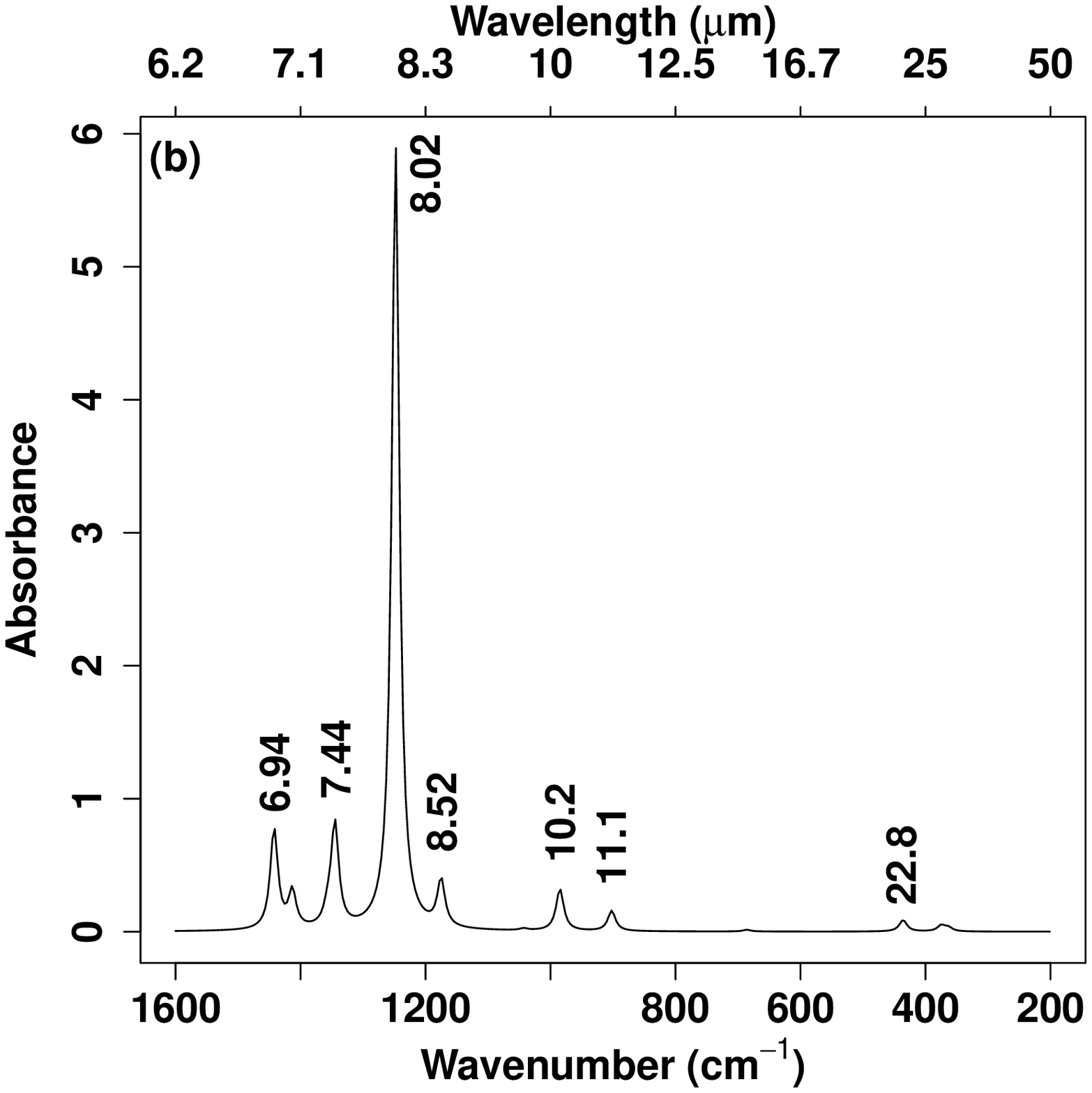}

\includegraphics[scale=0.4]{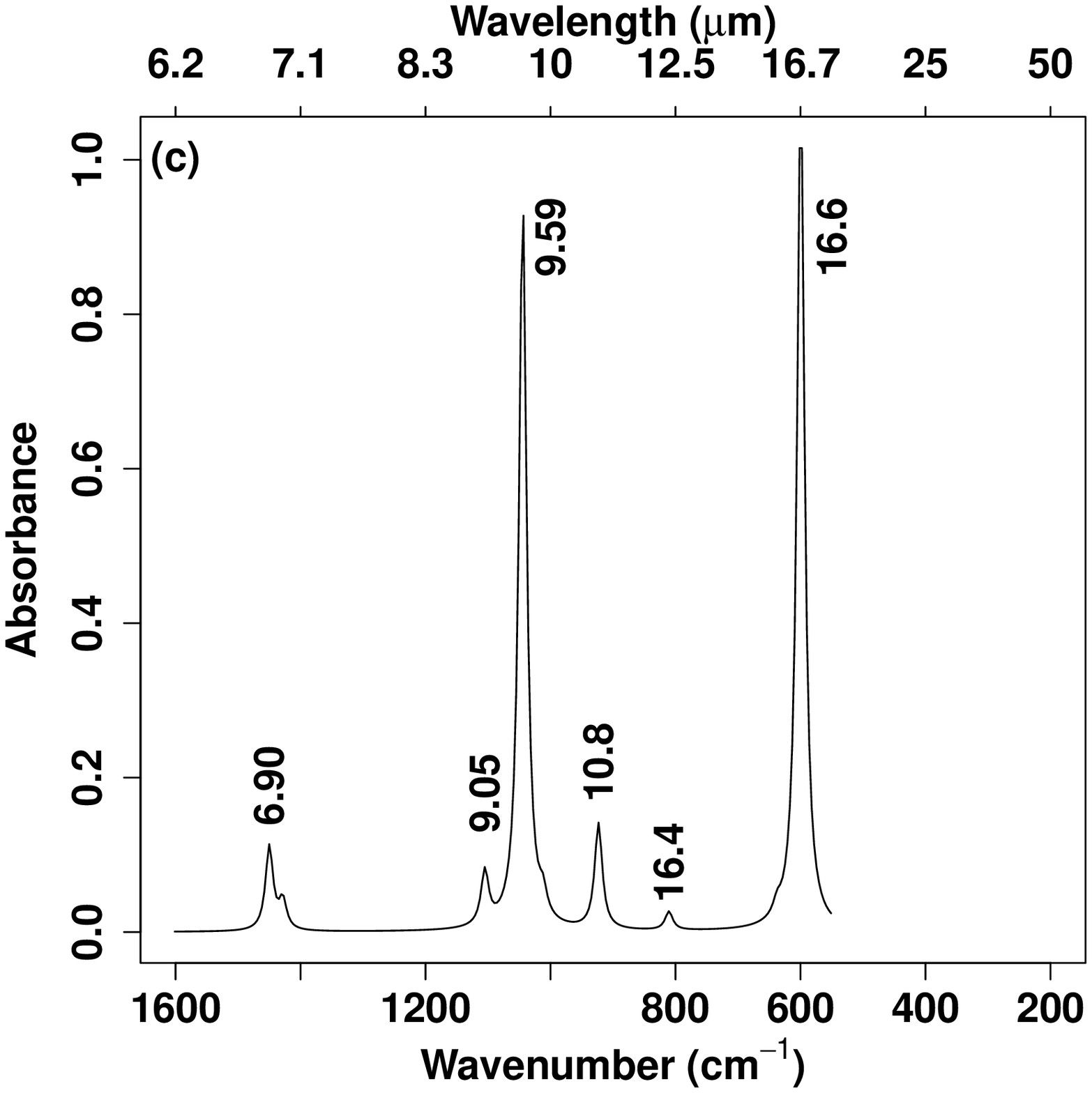}

\caption{IR spectrum  of:  (a) {\it trans}-HSSSH, 
(b) (CH$_3$)$_2$CS,
(c) C$_2$H$_4$S. Positions at the peaks are in $\mu$m.}
\label{fig6}
\end{figure}

\begin{table*}
\centering
\caption{Calculated  (scaled by 0.9687) and measured $^a$\citep{Klaboe1969,Ellestad1972}
vibrational fundamentals for 1,3,5 trithiane, (CH$_2$S)$_3$. After IR measurements in vapour $^a$\citep{Klaboe1969,Ellestad1972} we labeled relative intensities of experimental bands with: s (strong), m (medium), w (weak) and  v (very).  Three IR inactive A$_2$ modes (from 8 to 10) are omitted.}
\label{tab1}
  \begin{tabular}{ l| l|l|l| l|l| l| l|l|}
  \hline
  & Calculated & & & & Measured $^a$& & \\
  \hline
  Mode  &   Symm.  & $\widetilde{\nu}$ (cm$^{-1}$)  & $\lambda$ ($\mu$m)  &  IR int. (km mol$^{-1}$)  & Rel. int. &  $\widetilde{\nu}$ (cm$^{-1}$)      &  $\lambda$ ($\mu$m)  &  Rel. int. \\ 
  \hline 
  1  &   A$_1$  &  3027.2    & 3.30   &  1.08     & 3  &  -   &  -     & -           \\  
  2  &     &  2947.9    & 3.39   &  31.03    & 99  &  2900   &  3.45     & vs            \\   
  3  &     &  1392.5    & 7.18   &  8.95    & 28  &  1378   &  7.26     & s             \\   
  4  &     &  888.4     &  11.26  &  9.77    & 31   &   898  &  11.14     & s            \\   
  5  &     &  622.4     & 16.07   &   2.19   & 7 &  655    & 15.27      & s           \\   
  6  &     &  381.2    &  26.23  &   0.02    &  0&  398     &  25.13     & vw          \\   
  7  &     &  274.8   &  36.39  &   1.56    & 5 &  308     &   32.47    & m            \\    
    \hline    
 11   &   E  & 3028.7    &  3.30  &  0.02     & 0 & 2980          &  3.36    & s                \\     
 12   &     & 2953.4    &  3.39  &  3.95     & 13 & -          &  -   &  -                \\     
 13   &     & 1379.4    &  7.25  &  2.85     & 9 & 1394          &  7.17    & vs             \\     
 14   &     & 1204.5   &  8.30  &  13.31    & 42 &  1211          &  8.26    & vs                \\     
 15   &     & 1151.7    &  8.68  &  13.84     & 44 & 1162          &  8.61    & vs                 \\     
 16   &     &  769.6    & 13.00  &  0.04     & 0 & 802           &  12.47   & w                \\ 
 17   &     & 691.3    & 14.46  &  31.48    &  100 & 728          &  13.74   & vs                \\ 
 18   &     & 629.2     & 15.89 & 3.01      & 10 & 655           &  15.27   & s                  \\
 19   &     & 268.3     & 37.27  & 1.60     & 5 & 280           &  35.71   & m                 \\
 20   &     & 169.4    & 59.03  & 0.00      & 0 & -           &  -     & -                  \\
    \hline
  \end{tabular} 
\end{table*}

\begin{table*}
\centering
\caption{Calculated  (scaled by 0.9687) and measured $^a$\citep{Nash1988} vibrational A$_u$ fundamentals for 1,3,5,7-tetrathiocane,
(CH$_2$S)$_4$. Twenty one IR inactive A$_g$ modes are omitted. Details for relative intensities of measured bands as in Table 1.}
\label{tab2}
  \begin{tabular}{ l| l|l|l| l|l| l| l|l|}
  \hline
  & & Calculated & & &&  Measured $^a$ & & \\
  \hline
    Mode  & Symm. & $\widetilde{\nu}$ (cm$^{-1}$)  & $\lambda$  ($\mu$m)  &  IR int. (km mol$^{-1}$)  & Rel. int. &  $\widetilde{\nu}$ (cm$^{-1}$)      &  $\lambda$  ($\mu$m)  &  Rel. int. \\ 
  \hline 
   1 & A$_u$ &  3035.9    & 3.29   &  3.92     & 6  &  -   &  -     & -          \\  
   2 &  &  3031.5    & 3.30   &  2.72   & 4 &  -   &  -     & -            \\   
   3 &  &  2980.7    & 3.35   &  7.17    & 11  &  -   &  -     & -            \\   
   4 &  &  2978.1     & 3.36  &  14.39    & 23   &  -  &  -     & -            \\   
   5 &  &  1400.1     & 7.14   & 26.34   & 42 &  1397    & 7.16      & s           \\   
   6 &  &  1376.0    &  7.27  &   5.51    & 9&  1360     &  7.35     & m         \\   
   7 &  &  1210.7   &  8.26  &   49.84    & 79 &  1212     &   8.25    & m            \\      
   8 &  &  1203.7    &  8.31  &  29.81     & 47 & 1207          &  8.29    & s                \\     
   9 &  &  1136.6    &  8.80 &  6.93     & 11 & 1137          &  8.80   &  vw                \\     
  10 &  & 1122.2    &  8.91  &  6.89     & 11 &     -      &  -    & -            \\     
  11 &  & 797.4   &  12.54  &  2.22    & 4 &  804          &  12.44    & m                \\     
  12 &   & 789.7   &  12.66  &  19.86     & 31 & 750          &  13.33    & vs                 \\     
  13 &   & 716.6    & 13.96  &  28.23     & 45 & 732          &  13.66   & vs                \\ 
  14 &   & 692.8    & 14.43  &  63.27    &  100 & 715          &  13.99   & m                \\ 
  15 &   & 610.6     & 16.38 & 8.40      & 13 & 631           &  15.85   & w                  \\
  16 &   & 599.4     & 16.69  & 7.67     & 12 & 614           &  16.29   & vw                \\
  17 &   & 365.3    & 27.37  & 0.61      & 1 & -           &  -     & -                  \\
  18 &   & 278.5    & 35.91  & 7.56      & 12 & -           &  -     & -                  \\
  19 &   & 195.2    & 51.23  & 1.65      & 3 & -           &  -     & -                  \\
  20 &   & 128.1   & 78.07  & 3.55      & 6 & -           &  -     & -                  \\
  21 &   & 31.1    & 321.71  & 4.21      & 7 & -           &  -     & -                  \\
    \hline    
  \end{tabular} 
\end{table*}

\begin{table*}
\centering
\caption{Calculated  (scaled by 0.9687) vibrational fundamentals bands for pentathian, S$_5$CH$_2$.}
\label{tab3}
  \begin{tabular}{ l| l|l|l| l|}
  \hline
    Mode  & $\widetilde{\nu}$   (cm$^{-1}$)  & $\lambda$ ($\mu$m)  &  IR int. (km mol$^{-1}$)  & Rel. int.  \\ 
  \hline 
   1 & 3024.4  &  3.31    & 0.39   &   4               \\  
   2 & 2960.1  &  3.38    & 8.75   &   86            \\   
   3 & 1390.7  &  7.19    & 1.43   &   14                \\   
   4 & 1175.6  &  8.51     & 2.06   &  20                 \\   
   5 & 1084.8  &  9.22     & 0.12   &  1             \\   
   6 &  813.0 &  12.30    &  2.58  &  25           \\   
   7 &  670.2 &  14.92   & 1.18   &   12                \\      
   8 &  599.4 &  16.68    & 1.57   &  16                    \\     
   9 &  486.0 &  20.58    & 2.57  &   25                    \\     
  10 &  474.4 &  21.08   & 0.43   &   4               \\     
  11 &  417.9 &  23.93  & 10.13   &  100                   \\     
  12 &  357.6 &  27.96  &  0.26  &   3                      \\     
  13 &  342.5 &  29.19   & 2.07  &  20                     \\ 
  14 &  241.5 &  41.41   & 0.08  &  1                   \\ 
  15 &  222.0 &  45.05    & 0.09 &   1                     \\
  16 &  189.9 &  52.65    & 0.58  &  6                    \\
  17 &  173.2 &  57.72   & 2.35  &   23                    \\
  18 &  35.2 &  284.29   & 0.60  &    6                    \\
 
    \hline
  \end{tabular} 
\end{table*}

\begin{table*}
\centering
\caption{Calculated  (scaled by 0.9687) vibrational fundamentals bands for hexathiepan, S$_6$CH$_2$.}
\label{tab4}
  \begin{tabular}{ l| l|l|l| l|}
  \hline
    Mode  & $\widetilde{\nu}$  (cm$^{-1}$)  & $\lambda$  ($\mu$m)  &  IR int. (km mol$^{-1}$)  & Rel. int.  \\ 
  \hline 
   1 & 3006.8  &  3.33    & 2.29   &  36                \\  
   2 & 2951.7  &  3.39    & 2.37   &  37             \\   
   3 & 1386.2  &  7.21    & 1.38   &  22                 \\   
   4 & 1175.2  &  8.51     & 4.78   & 75                  \\   
   5 & 1094.9  &  9.13     & 0.02   &  0             \\   
   6 & 790.8  &  12.65    & 3.21   &  50           \\   
   7 & 688.0  &  14.53   &  2.87  &   45                \\      
   8 & 591.3  &  16.91    & 0.01   &   0                   \\     
   9 & 460.5  &  21.71    & 1.46  &  23                     \\     
  10 & 456.6  &  21.90   & 1.44   &  23               \\     
  11 & 448.3  &  22.31  & 6.41   &  100                   \\     
  12 & 403.0  &  24.81  & 0.52   &   8                     \\     
  13 & 373.0  &  26.81   & 1.08  &  17                     \\ 
  14 & 347.3  &  28.79   & 0.00  &   0                  \\ 
  15 & 270.6  &  36.96    & 0.47 &   7                    \\
  16 & 228.7  &  43.73    & 1.07  & 17                    \\
  17 & 193.0  &  51.81   & 4.17  &  65                     \\
  18 & 174.8  &  57.21   & 0.07  &   1                    \\
 19 &  161.3  &  62.01   & 0.53  &   8                   \\
 20 &  111.0  &  90.07   & 0.23  &   4                     \\
 21 &   43.5 &  229.67   & 2.03  &   32                    \\
 
    \hline
  \end{tabular} 
\end{table*}

\begin{table*}
\centering
\caption{Calculated  (scaled by 0.9687) and measured $^a$\citep{Wieser1969} fundamental  vibrational bands for HSSSH. Relative intensities of experimental bands are labeled with: sh (shoulder), m (medium), and strong (s).}
\label{tab5}
  \begin{tabular}{ l| l|l|l| l|l|l|l|l|}
  \hline
  & & Calculated & & & & Measured $^a$ &  &\\
  \hline
    Mode  & Symm.& $\widetilde{\nu}$ (cm$^{-1}$)  & $\lambda$  ($\mu$m)  &  IR int. (km mol$^{-1}$)  & Rel. int. & $\widetilde{\nu}$ (cm$^{-1}$) & $\lambda$  ($\mu$m) & Rel. int.\\ 
  \hline 
   1 & A  & 2548.6     & 3.92   &  0.48    &    2   & 2540   & 3.94  &  sh  \\  
   3 &   & 841.9     & 11.88   &  0.04     &   0   & 868   &  11.52 &  sh  \\   
   5 &  &  458.8    &  21.79   &  0.10     &   0   & 487   &  20.53 &  sh   \\   
   8 &   & 303.4      &32.96    & 17.20     &  58  &  -  &  -       &  -   \\   
   9 &    & 193.7      & 51.63   & 0.00      &  0  &  -  &  -       &  -   \\   
   \hline
   2 & B &  2547.1    & 3.93   &   0.90    &   3   &  2532   & 3.95  & m    \\   
   4 &   &  829.2   &  12.06  &   8.54     &  29  &  856   & 11.68  & m     \\      
   6 &   &  438.2    & 22.82   &  29.67     & 100 &  477   & 20.96  & s     \\   
   7 &   &  322.2   &  31.04  &  11.77     &   40 &   -  &   -      &   -  \\      
    \hline
  \end{tabular} 
\end{table*}

\begin{table*}
\caption{Calculated  (scaled by 0.9687) and measured vibrational fundamental bands for thioacetone, (CH$_3$)$_2$CS.  Measured bands are from 
$^a$\citep{Lipscomb1970} and $^b$\citep{NIST2014} 
in vapour, as well as  from $^c$\citep{Garrigou1974} in liquid. Relative intensities of experimental bands are labeled with:  m (medium), s (strong), and vs (very strong). Raman measurements in liquid are labeled by r. }
\label{tab6}
  \begin{tabular}{ l| l|l|l| l|l|l|l|l|l|l|}
  \hline
          &      & Calculated & & & & Measured $^{a,b}$ &  & Measured $^c$ & &\\
  \hline
    Mode  & Symm.& $\widetilde{\nu}$ (cm$^{-1}$)  & $\lambda$  ($\mu$m)  &  IR int. (km mol$^{-1}$)  & Rel. int.  & $\widetilde{\nu}$ (cm$^{-1}$) & $\lambda$  ($\mu$m) & $\widetilde{\nu}$ (cm$^{-1}$) & $\lambda$  ($\mu$m) &  Rel. int.\\ 
  \hline 
   1 & A$_1$  & 3042.9    & 3.29   &  2.22      &    2   & -&   - &- & -& -\\  
   2 &   & 2920.6     & 3.42   &  21.43     &   15   &- & - & -& -&-\\   
   3 &   & 1446.4     & 6.91    &  0.15     &   0   &- &   - & 1423& 7.03 & m\\   
   4 &   & 1356.7      & 7.37   &  2.46     &  2    &- &    -  &1359 & 7.36 & r\\   
   5 &   & 1247.9      & 8.01   & 139.27    &  100  &1274  $^a$ &  7.85 &1269 & 7.88 &vs\\   
   6 &  &   985.0   & 10.15   &  7.46     &  5     & -&-  & -& -&-\\   
   7 &   & 685.8    & 14.58   &  0.30      &  0    &- &    -   & 704& 14.20& r\\      
   8 &   & 363.5     & 27.51   & 0.69      & 0     & 363  $^b$ &  27.548  & 380& 26.32 &s\\  
   \hline 
   9 & A$_2$  & 2953.0 &  3.39     &  0       &    -       &    -   &  -    & - & - &- \\
   10 &   & 1420.1 & 7.04      &  0       &   -        &   -    &  -   & 1447 & 6.91 & m \\
   11 &   & 895.3 &  11.17     &  0       &   -        &   -    &  -   & 992 & 10.08& m \\
   12 &   & 79.1 &   126.43    &  0       &   -        &   -    &  -    & - &- & -\\ 
   \hline
   13 & B$_1$  & 2960.0    & 3.38   & 15.87      &  11  &- &    -   & -& - &-\\  
   14 &   & 1442.3    & 6.93   & 17.78      & 13   &- &   -    & 1438& 6.95 & r\\     
   15 &   & 1042.9    & 9.59   &  0.34     &  0  &- &   -     & 1087& 9.20  &vs\\     
   16 &   &  435.9   & 22.94   &  2.04     &  1  &- &    -    &436 & 22.94 &m\\  
   17 &   &  152.6   & 65.53   &  0.50     &  0  &153  $^b$&  65.359   & -&- &-\\ 
   \hline 
   18 & B$_2$  &  3041.2   & 3.29   &  8.59     &  6  &- &   -   & - & -&-\\  
   19 &   &  2912.5   & 3.43   &  1.57     &  1  &- &   -   & -& -&-\\     
   20 &   &  1413.2   &  7.08  &  6.56     &  5  &-&    -   &1423 & 7.03 &m\\     
   21 &   &  1345.0   & 7.44   & 18.47     &  13  &- &  -   &1353  & 7.39&m\\  
   22 &   &  1175.1   & 8.51   &  8.36     &  6  &- &   -   &1195 & 8.37&m\\ 
   23 &   &  902.2   & 11.08   &  3.61     &  3  &- &   -   & 896&11.16& r\\    
   23 &   &   374.7  & 26.68   &  1.06     & 1   &- &   -   &- & -&-\\  
            
    \hline

  \end{tabular} 
\end{table*}

\begin{table*}
\centering
\caption{Calculated  (scaled by 0.9687) and measured $^a$\citep{Allen1986}
vibrational fundamental bands for thiirane, C$_2$H$_4$S. Three IR inactive A$_2$ modes are omitted.}
\label{tab7}
  \begin{tabular}{ l| l|l|l| l|l|l|l|}
  \hline
          &      & Calculated & & & & Measured$^a$ &  \\
  \hline
    Mode  & Symm.& $\widetilde{\nu}$ (cm$^{-1}$)  & $\lambda$  ($\mu$m)  &  IR int. (km mol$^{-1}$)  & Rel. int.  &  $\widetilde{\nu}$(cm$^{-1})$ & $\lambda$  ($\mu$m)\\ 
  \hline 
   1 & A$_1$  & 3027.5     & 3.30   &  15.89     &  62  & 3014& 3.32   \\  
   2 &   & 1449.9     & 6.90   &   2.58    &   10   & 1457& 6.86 \\   
   3 &   & 1105.0     & 9.05    &  1.66     &   6   &1109 & 9.02     \\   
   4 &   & 1012.0      & 9.88   &  0.68     &  3    &1024 & 9.76       \\   
   5 &   &  599.6     & 16.68   & 25.59    & 100   & 627&  15.95  \\  
   \hline 
   9 & B$_1$  &  3115.5    & 3.21   &  3.36     &  13     & 3088& 3.24\\   
   10 &   &  923.3   &  10.83  &  3.24      & 13     & 945&  10.58     \\      
   11 &   &  810.5    &  12.34  & 0.57      & 2     &  824&  12.14  \\ 
   \hline  
   12 & B$_2$  & 3026.7    & 3.30   & 11.57      & 45   & 3013&   3.32    \\  
   13 &   & 1428.5    & 7.00   &  0.90     &   4   & 1436&   6.96    \\     
   14 &   & 1044.4    & 9.57   &  22.33     &  87  &1051 &   9.51    \\     
   15 &   &  637.6   & 15.68   &   0.38    &  1  & - &     -  \\
    \hline
  \end{tabular} 
\end{table*}

Our results for
IR spectra are shown in Figs. 4-6, whereas the bands and their intensities are presented in Tables~\ref{tab1}-\ref{tab7}.
Relative intensities of calculated bands shown in these Tables are determined as percentage of the strongest band. Relative intensities of measured bands are known only qualitatively (v--very, s--strong, m--mean, w--weak, e--extremely), and were after \citep{Klaboe1969,Ellestad1972,Wieser1969}
assigned a quantitative measure by dividing the interval from 0 to 1 into six equal parts.

In (CH$_2$S)$_3$ the strongest calculated band at  14.46 $\mu$m is dominantly due to the C-S stretching (Table~\ref{tab1}). 
The same is valid for the pair of the (CH$_2$S)$_4$ calculated rather strong bands at 12.66 and 13.96 $\mu$m 
 (Table~\ref{tab2}). The strongest calculated band of (CH$_2$S)$_4$ 
 is at 14.43 $\mu$m.
We are not aware of any existing measured IR vibrational data concerning pentathian and hexathiepan.  
The vibrational normal modes involving sulfur atoms are generally weak. This is particularly well seen in pentathian and hexathiepan where the strongest bands, respectively at 23.93 and 22.31 $\mu$m, 
are only 10.13 and 6.41 km mol$^{-1}$ strong 
(Tables~\ref{tab3} and \ref{tab4}).
 
In the case of HSSSH, the strong band calculated at 22.82   $\mu$m 
is the antisymmetric S-S stretching (Table~\ref{tab5}). Vibrational spectra of H$_2$S$_3$ were first measured in CCl$_4$ and CS$_2$ solutions, but the {\it cis} conformer was assigned as a carrier
\citep{Wieser1969}. More recently, the IR spectrum of H$_2$S$_3$ was measured in the gas phase in the range from 400 to 4000 cm$^{-1}$, but unfortunately with a low resolution, and only frequencies measured at  2548, 865, and 480 cm$^{-1}$ were reported
\citep{Liedtke1993}. 
However, both measurements \citep{Wieser1969,Liedtke1993} and our results show that IR frequencies of {\it trans} and {\it cis} conformers are similar. In Table~\ref{tab5} we compare our results with detailed measurements \citep{Wieser1969}. 
Liedtke and coworkers also reported the calculated IR frequencies done using the M{\/o}ller-Plesset perturbation operator (MP2) and the  MC-311G(d,p) basis set. Their frequencies are unscaled and bands start  
at much higher (for 243 cm$^{-1}$) values than experimental and these we calculated.

The only band of thioacetone worth of searching for in space should be a very strong C=S stretching band calculated at 8.01 $\mu$m (Table~\ref{tab6}).
Thiirane can be detected by measuring its two strongest bands calculated  at 16.68 and  9.57 $\mu$m 
 (Table~\ref{tab7}). The former can be described as a symmetric C-S stretching, while the latter as a CH$_2$ wagging.
First  fourteen modes in the IR spectra of thiirane were measured and assigned by \citep{Allen1986}. Recently, far-IR modes at 760--400 and 
170--10 cm$^{-1}$ were studied using the synchrotron light source \citep{Bane2012}. Two modes $\nu _5=628.1$ and $\nu _{15}=669.7$ cm$^{-1}$ were
measured, as well as subsequently calculated at the B3LYP/cc-pVTZ level, and then fitted. 

For all molecules we study some other rather strong bands exist, but they are typical for the C-H bonds. Thus, these bands are less suitable for a search of the sulfur-bearing species in space.  

In Table~\ref{tab8} we compare calculated bands of (CH$_2$S)$_3$, (CH$_2$S)$_4$, and C$_2$H$_4$S
with unidentified features observed by {\it ISO} in, either galaxies known  for their starburst activity, or galaxies with active galactic nuclei: M82, NGC 253, the 30 Doradus star-forming region in the Large Magellanic Cloud, NGC 1068, Circinus, Arp 220 \citep{UnidentifiedISO,Sturm2000,Lutz2000,Fischer1999}.
We find that in the interval around unidentified bands, determined by the scaling factor used within density functional theory method, belong ten calculated bands of (CH$_2$S)$_3$, eleven of (CH$_2$S)$_4$, and five of C$_2$H$_4$S. For all three species two bands with highest intensities (labeled bold in Table~\ref{tab8}) belong to the interval around unidentified bands.
Therefore, (CH$_2$S)$_3$, (CH$_2$S)$_4$, and C$_2$H$_4$S are good candidates for carriers of several unidentified bands observed by {\it ISO} \citep{UnidentifiedISO,Sturm2000,Lutz2000,Fischer1999}. 
{\it ISO} observations showed that intensities of bands vary considerably depending on the extragalactic source \citep{Sturm2000}. The strongest calculated bands in Table 8 are at 3.3, 3.4, 8.3, 14.8, and 16.5 $\mu$m. These positions agree with bands in some of objects observed by {\it ISO}. The band at 3.3 and its satellite at 3.4 $\mu$m are strong in M82, NG253, 30 Doradus, and  Circinus. This feature is weaker in NGC 1068.  The band at 8.3 $\mu$m  was observed only in M82, and it is not strong. The band at 14.8 $\mu$m is also not strong. It was detected in M82, NGC253, 30 Doradus, and Circinus. The band at 16.5 $\mu$m was observed in three galaxies. It is strong in M82 and NGC 253, whereas it is weak in the 30 Doradus region. Differences of intensities in observed spectra are results of different physical conditions in extragalactic objects.

Table~\ref{tab9} shows that
in the proposed interval around  unidentified bands are positioned five bands of S$_5$CH$_2$, nine of S$_6$CH$_2$, three of HSSSH, and ten of (CH$_3$)$_2$CS. However, either two strongest bands (S$_5$CH$_2$, (CH$_3$)$_2$CS), or only the strongest one (S$_6$CH$_2$, HSSSH) do not belong to this interval. Because of high quality basis set that has been shown to produce accurate IR
intensities \citep{Zvereva2011}, S$_5$CH$_2$, S$_6$CH$_2$, HSSSH, and (CH$_3$)$_2$CS are less probable  carriers of unidentified features observed in selected galactic sources by
{\it ISO} \citep{UnidentifiedISO,Sturm2000,Lutz2000,Fischer1999}.

\begin{table*}
\caption{Comparison of unidentified lines and bands in the (2--200) $\mu$m range observed by {\it Infrared Space Observatory} $^a$\citep{UnidentifiedISO,Sturm2000,Lutz2000,Fischer1999}, and our results for the trimer and tetramer of thioformaldehyde, as well as for thiirane. The two strongest calculated bands are shown in bold and $\Delta = 18.9$ cm$^{-1}$ is the mean average error after using the scaling factor within the density  functional theory calculation method.}
\label{tab8}
  \begin{tabular}{ l|l|l| l|l|l|l|}
  \hline
                                    &                         &                             &             &    (CH$_2$S)$_3$               & (CH$_2$S)$_4$ &C$_2$H$_4$S  \\
  \hline
    $\lambda$  ($\mu$m)$^a$ & $\widetilde{\nu}$ (cm$^{-1}$)$^a$  & $\widetilde{\nu}$$- \Delta$ (cm$^{-1}$) &  $\widetilde{\nu}$$+ \Delta$ (cm$^{-1}$) &  $\widetilde{\nu}$ (cm$^{-1}$) &  $\widetilde{\nu}$ (cm$^{-1}$)  &  $\widetilde{\nu}$ (cm$^{-1}$)   \\ 
  \hline 
   3.25  & 3076.92   & 3058.0   & 3095.8    &   -                   &    -            & -                    \\  
   -     &  -        & -         & -          &   -                    &    -           & 3115.5               \\
   3.3   & 3030.30   & 3011.4   & 3049.2    & 3028.7; 3027.2         & 3035.9; 3031.5  & {\bf 3027.5}; 3026.7 \\  
   -     & -         &  -        & -          &  -                    & 2980.7          &  -                    \\
   -     &  -        &  -        & -          &  -                    & 2978.1          &  -                    \\      
   3.4   & 2941.18   & 2922.3   & 2960.1    & 2953.4; {\bf  2947.9} &    -            &  -                    \\   
   3.5   & 2857.14   & 2838.2   & 2876.0    & -                     &    -            & -                     \\   
   3.75  & 2666.67   & 2647.8   & 2685.6    & -                     &    -            & -                      \\   
   5.25  & 1904.76   & 1885.9   & 1923.7    & -                     &    -            &-                        \\      
   5.65  & 1769.91   & 1751.0   & 1788.8    & -                     &  -              & -                       \\ 
   6.0   & 1666.67   & 1647.8   & 1685.6    & -                     &  -              & -                      \\
   6.2   & 1612.90   & 1594.0   & 1631.8    &   -                  &  -              & -                       \\
   6.3   & 1587.30   & 1568.4   & 1606.2    &   -                  &  -              & -                       \\
   -     & -         & -         & -          & -                    & -               & 1449.9                  \\
   7.0   & 1428.57   & 1409.7   & 1447.5    &   -                  &  -              & 1428.5                  \\ 
   -     & -         & -         & -          & -                     &  1400.1         &  -                      \\
   -     & -         & -         & -          & 1392.5; 1379.4       &  1376.0           & -                        \\      
   7.555 & 1323.63   & 1304.7   & 1342.5    &  -                    &-                & -                       \\  
   7.6   & 1315.79   & 1296.9   & 1334.7    &  -                    &-                & -                    \\     
   7.8   & 1282.05   & 1263.2   & 1301.0    &  -                    &-                & -                  \\     
   8.3   & 1204.82   & 1185.9   & 1223.7    & 1204.5               &{\bf 1210.7}; 1203.7 &  -                 \\  
   8.6   & 1162.79   & 1153.9   & 1181.7    & -                     & -                   & -               \\ 
   -     & -         & -         &  -         & 1151.7                & -                  & -                \\
   -     & -         &  -        &  -         &  -                    & 1136.6             & -  \\
   -     & -         &  -        &  -         & -                     & 1122.2             & -   \\
   -     & -         &  -        &  -         &  -                    &  -                  &  1105.0   \\
   -     & -         &  -        &  -         & -                     & -                   &  1044.4    \\
   -     & -         &  -        &  -         & -                     & -                   &  1012.0    \\ 
   10.6  &  943.40   & 924.5    & 962.3     & -                     &   -                &  -\\  
   -     &  -        & -         & -          & -                     & -                  & 923.3 \\
   11.05 &  904.98   & 886.1    & 923.9     & -                     &   -                & -\\     
   11.25 &  888.89   & 870.0    & 907.8     & 888.4                &    -               & -  \\     
   12.0  &  833.33   & 814.4    & 852.2     &  -                    &  -                 & -  \\  
   -     &  -        &  -        & -          & -                     &                    & 810.5       \\
   12.7  &  787.40   & 768.5    & 806.3     & 769.6                & 797.4; 789.7       & -   \\ 
   13.55 &  738.01   & 729.1    & 756.9     & -                     &  -               &-  \\  
   -     &  -        & -         & -          & -                     & 716.6           & - \\
   14.25 &  701.75   & 682.8    & 720.6     & -                     &   -                &- \\  
   14.8  &  675.68   & 656.8    & 694.6     &{\bf 691.3}           & {\bf 692.8}        & -         \\
   15.7  &  636.94   & 618.0    & 655.8     & 629.2; 622.4                &  -                  &  637.6          \\ 
   16.5  &  606.06   & 587.2    & 625.0     &  -                    & 610.6; 599.4       & {\bf 599.6}         \\
   17.4  &  574.71   & 555.8    & 593.6     &  -                    & -                   & -            \\
   18.0  &  555.56   & 536.7    & 574.5     & -                     & -                  &  -          \\ 
   -     & -         & -         & -          & 381.2                & -                   & -           \\
   -     &  -        & -         & -          &  -                    & 365.3              & -           \\
   34.0  &  294.12   & 275.2    & 313.0     & -                     & 278.5              & -           \\
   -     &  -        & -         & -          & 274.8                 & -                  & -           \\
   -     & -         & -         & -          & 268.3                & -                   & -           \\
    -    & -         & -         & -          &  -                    & 195.2              & -          \\ 
   60.12 &  166.33   & 147.4    & 185.2     & -                     &  -                  & -            \\    
   74.24 &  134.70   & 119.5    & 153.6     & -                     & 128.1                 & -           \\
    -    &  -        & -         & -          & 169.4                &  -             & -           \\
77.155   &  129.61   & 110.7    & 148.5     &  -                    & -                   & -           \\
153.12   &   65.31   & 46.4     & 84.2      &  -                    & -                   & -           \\   
     -   &  -        & -         & -          & -                    &  31.3              &  -    \\    
    \hline
  \end{tabular} 
\end{table*}

\begin{table*}
\caption{Comparison of unidentified lines and bands in the (2--200) $\mu$m range observed by {\it Infrared Space Observatory} $^a$\citep{UnidentifiedISO,Sturm2000,Lutz2000,Fischer1999}, and our results for pentathian, hexathiepan,  trisulfane, and thioacetone. The two strongest calculated bands are shown in bold and $\Delta = 18.9$ cm$^{-1}$ is the mean average error after using the scaling factor within the density  functional theory calculation method.}
\label{tab9}
  \begin{tabular}{ l|l|l| l|l|l|l|l|}
  \hline
                                   &                         &                             &             &    S$_5$CH$_2$   & S$_6$CH$_2$ & HSSSH& (CH$_3$)$_2$CS\\  
  \hline
    $\lambda$  ($\mu$m)$^a$ & $\widetilde{\nu}$ (cm$^{-1}$)$^a$  & $\widetilde{\nu}$$- \Delta$ (cm$^{-1}$) &  $\widetilde{\nu}$$+ \Delta$ (cm$^{-1}$) & $\widetilde{\nu}$ (cm$^{-1}$)   & $\widetilde{\nu}$ (cm$^{-1}$)    &$\widetilde{\nu}$ (cm$^{-1}$) & $\widetilde{\nu}$ (cm$^{-1}$) \\ 
  \hline 
   3.25  & 3076.92   & 3058.0   & 3095.8    &  -                   &  -              &   -          &-\\       
   3.3   & 3030.30   & 3011.4   & 3049.2    &  3024.4              &  -              &  -           &3042.9; 3041.2\\  
   -     & -         &  -        & -          &   -                  & 3006.8          &  -           &-\\
   -     &  -        &  -        & -          &  {\bf 2960.1}        & -               &  -           &-\\      
   3.4   & 2941.18   & 2922.3   & 2960.1    &  -                   & 2951.7          & -            &2960.0\\
   -     &  -        & -         & -          &  -                   & -               & -            & {\bf 2920.6}; 2912.5 \\   
   3.5   & 2857.14   & 2838.2   & 2876.0    &  -                   & -               &  -           &-\\   
   3.75  & 2666.67   & 2647.8   & 2685.6    &  -                   & -               & -            &-\\
    -    &  -        & -         & -          & -                    & -               & 2548.6       &- \\ 
    -    &  -        & -         & -          & -                    & -               & 2547.1       & -\\   
   5.25  & 1904.76   & 1885.9   & 1923.7    &  -                   & -               & -            &- \\      
   5.65  & 1769.91   & 1751.0   & 1788.8    &  -                   & -               &  -           &-\\ 
   6.0   & 1666.67   & 1647.8   & 1685.6    &  -                   & -               & -            &-\\
   6.2   & 1612.90   & 1594.0   & 1631.8    &  -                   & -               & -            &-\\
   6.3   & 1587.30   & 1568.4   & 1606.2    &  -                   & -               &  -           &-\\
   7.0   & 1428.57   & 1409.7   & 1447.5    &  -                   & -               & -            &1446.4; 1442.3; 1413.2\\ 
   -     & -         & -         & -          &  1390.7              & 1386.2          & -            &1356.7; 1345.0\\      
   7.555 & 1323.63   & 1304.7   & 1342.5    &  -                   &  -              & -            &-\\  
   7.6   & 1315.79   & 1296.9   & 1334.7   &  -                   & -               & -            &-\\     
   7.8   & 1282.05   & 1263.2   & 1301.0    &  -                   & -               & -            &-\\ 
   -     &  -        & -         & -          &  -                   & -               & -            & {\bf 1247.9} \\    
   8.3   & 1204.82   & 1185.9   & 1223.7    &  -                   & -               & -            &-\\  
   8.6   & 1162.79   & 1153.9   & 1181.7    &  1175.6              & {\bf 1175.2}    & -            &1175.1\\ 
   -     & -         & -         &  -         &  1084.8              & 1094.9          & -            &1042.9; 985.0\\
   10.6  &  943.40   & 924.5    & 962.3     &  -                   & -               & -             &-\\  
   11.05 &  904.98   & 886.1    & 923.9     &  -                   & -               &-              &902.2\\     
   11.25 &  888.89   & 870.0    & 907.8     &  -                   & -               & -             &- \\     
   12.0  &  833.33   & 814.4    & 852.2    &  -                   & -               &  841.9; 829.2 &-\\  
   -     &  -        &  -        & -          &  813.0               & -               &  -            &-\\
   12.7  &  787.40   & 768.5    & 806.3     &  -                   & 790.8           &  -            &-\\ 
   13.55 &  738.01   & 729.1    & 756.9     &  -                   & -               & -             &-\\  
   14.25 &  701.75   & 682.9    & 720.6     & -                    & 688.0           & -             &685.8\\  
   14.8  &  675.68   & 656.8    & 694.6     & 670.2                &  -              & -             &- \\
   15.7  &  636.94   & 618.0    & 655.8     &  -                   & -               & -             &-   \\ 
   16.5  &  606.06   & 587.2    & 625.0     & 599.4                & 591.3           & -             & -\\
   17.4  &  574.71   & 555.8    & 593.6     & -                    & -               & -             & -   \\
   18.0  &  555.56   & 536.7    & 574.5     & -                    & -               & -             &-   \\ 
   -     & -         & -         & -          & 486.0                & 460.5           & 458.8         & - \\
   -     &  -        & -         & -          & 474.4                & 456.6           & {\bf 438.2}   & -   \\
   -     & -         & -         & -          & {\bf 417.9}          & {\bf 448.3}     & -             & -  \\
   -     & -         & -         & -          & 357.6                & 403.0           & -             & -   \\
   -     & -         & -         & -          & 342.5                & 373.0           & -             & 435.9  \\
   -     & -         & -         & -          & -                    & 347.3           & 322.2         &374.7; 363.5\\
   34.0  &  294.12   & 275.2    & 313.0     & -                    &  -              & {\bf 303.4}   & -   \\
   -     &  -        & -         & -          & 241.5                &  270.6          &   -           &  \\
   -     & -         & -         & -          & 222.0                &  228.7          &   -           & - \\
    -    & -         & -         & -          & 189.9                &  193.0          &  193.7        & -    \\ 
   60.12 &  166.33   & 147.4    & 185.2     & 173.2                &  174.8; 161.3          &  -            & 152.6  \\  
   74.24 &  134.70   & 119.5    & 153.6     &  -                   &  -              &  -            &  -   \\
77.155   &  129.61   & 110.7    & 148.5     &  -                   & 111.0           &  -            & -    \\
153.12   &   65.31   & 46.4     & 84.2      &  -                   & -               &  -            & -   \\  
     -   &  -        & -         & -          &  35.2                &   -             &  -            & -\\    
    \hline
  \end{tabular} 
\end{table*}

\subsection{Sulfur-bearing species in extragalactic molecular gases}
\label{molgal}

Below we present the list of the presently detected (to the best of our knowledge) sulfur-bearing species in NGC 253, M82, NGC1068, Circinus, Arp 220, and 30 Doradus. These objects were observed by {\it ISO} and unidentified features were found 
\citep{UnidentifiedISO,Sturm2000,Lutz2000,Fischer1999}. Ultraviolet, X-ray, and cosmic-ray irradiations, as well as shocks, are present in these galaxies and determine the chemical processes in molecular gases for sulfur-bearing and other species. For example, it was found in laboratories on Earth that UV and X-ray irradiations catalyze the formation of (CH$_2$S)$_3$ and (CH$_2$S)$_4$ from H$_2$CO and H$_2$S \citep{Credali1967}. Therefore, favorable conditions for the formation of (CH$_2$S)$_n$ exist in extragalactic molecular gases.

NGC 253 is also known as the Sculptor, or the Silver dollar galaxy. 
It is the bright, spiral, starburst galaxy
where many molecular lines have been detected.
Sulfur-bearing species  H$_2$S, CS, C$_2$S, NS, SO, H$_2$CS, OCS, and SO$_2$  were observed  in 
the nuclear region of NGC 253  
using  the {\it IRAM} 30 m and the Swedish-ESO Submillimetre Telescope ({\it SEST})
\citep{Martin2005,Martin2006}.
It was proposed that turbulent motions and shocks release H$_2$S and OCS from grain mantles into the gas phase.
Then other sulfur-bearing species are formed in the gas-phase chemistry processes.
Relative abundances of sulfur-bearing molecules in different environments suggest
that the  chemical processes in the nuclear region of NGC 253  are similar to those in the Sgr B2
envelope in the galactic center \citep{Martin2008}. Low-velocity large scale shocks were found responsible for the similar sulfur chemistry
in these two regions. Both formaldehyde \citep{Gardner1974} and H$_2$S \citep{Martin2005,Martin2006}, which react in laboratories under irradiation and form (CH$_2$S)$_n$, were detected in NGC 253.

M82, also known as NGC 3034 or  Cigar, is an irregular starburst galaxy.
It was found that the chemical processes in M82 and NGC 253 are different
\citep{Martin2006, Aladro2011}. Although both galaxies are starburst, they
are at the different evolutionary stage. M82 is more evolved starburst where 
photodissociation regions  \citep{Tielens1985}
are dominant in the heating of the molecular gas.
These processes are driven by ultraviolet photons from young stars. 
Molecular abundances in M82 are similar to the well studied photon-dominated region
the Orion bar.
Photon-dominated regions are also present in NGC 253 \citep{Martin2009}, 
but they do not drive the heating of molecular clouds.
{\it IRAM} 30 m telescope was used to study the chemical species in the central parts of M82 \citep{Aladro2011}. 
Among eighteen detected molecules several sulfur species were found: H$_2$S, H$_2$CS, CS, SO, and SO$_2$.
Formaldehyde was also detected in M82 \citep{Aladro2011} and could under irradiation react with H$_2$S to form (CH$_2$S)$_n$.

NGC 1068, also known as M77,  is a barred, spiral, active galaxy.
Although NGC 1068 is with an active  galactic nucleus, it also contains the starburst ring
\citep{Takano2014,Garcia2014,Viti2014}.  
The chemical composition  of molecular clouds in NGC 1068 was studied by 
the {\it IRAM} 30 m telescope \citep{Aladro2013}. 
Seventeen molecules and seven isotopologues were 
detected and among them sulfur species SO, NS, and CS.
The results for NGC 1068 were compared to the starburst galaxies M82 and NGC 253.
Some sulfur molecules which were detected in two starburst galaxies, such as OCS, SO$_2$, 
C$_2$S, H$_2$S, and H$_2$CS, have not been yet found in NGC 1068.
Recently NGC 1068 was studied using the Atacama Large Millimeter Array ({\it ALMA}) 
\citep{Takano2014,Garcia2014,Viti2014}.  Takano and coworkers studied several molecular transitions
(including those in sulfur species CS and SO) in the central region of NGC 1068. 
Garcia-Burillo and coworkers \citep{Garcia2014},  as well as Viti and coworkers \citep{Viti2014} also investigated the molecular gas in NGC 1068 using the {\it ALMA} telescope.
They mapped the emission of several tracers (including CS) and underlying continuum emission
of the circumnuclear disk and the starburst ring.

The Circinus galaxy is  an active, starburst, spiral galaxy.   NGC 1068 and
Circinus are similar. Both are typical  Seyfert 2 galaxies close to the Milky Way
and contain the molecular gas which surrounds nuclei, as well as star-forming rings
\citep{Zhang2014}.
Several molecular transitions, and among them
in sulfur species CS and SO, were observed in the study of 
a dense molecular gas in the central part of the Circinus galaxy 
using the {\it SEST} telescope 
\citep{Curran2001}.

Arp 220 is an ultra-luminous infrared galaxy (ULIRG)  closest to Earth. 
This elliptical galaxy is the merger system and  contains a star-forming region
as well as a double nucleus.  The prebiotic molecule methanimine  (CH$_2$NH) was detected in Arp 220
\citep{Salter2008}. The molecular composition of Arp 220 was studied using  the Submillimeter Array
({\it SMA}) in Mauna Kea, Hawaii \citep{Martin2011}.
Fifteen molecular species and six isotopologues were observed, and among them
sulfur-bearing molecules: H$_2$S, SO, NS, and CS. Formaldehyde was also detected.

The star-forming region 30 Doradus in the Large Magellanic cloud,  also known as the Tarantula nebula or NGC 2070,
is the most active in the Local group. Heikkila and coworkers studied molecular abundances in
30 Doradus using the {\it SEST} telescope \citep{Heikkila1999}.  They detected
several molecular transitions and among them in CS and SO. The sulfur-bearing molecule
CS in 30 Doradus was also observed   by Rubio and coworkers using {\it SEST}
\citep{Rubio2009}, and 
in recent studies by the {\it ALMA} telescope \citep{Indebetouw2013}. 

We find that all six bright bands from Table 8 agree with features observed in M82.
Five strong calculated bands agree with observed features in NGC 253 and 30 Doradus, three in Circinus,
and only one in NGC 1068. Therefore, starburst galaxies are more favorable for the formation
of (CH$_2$S)$_3$, (CH$_2$S)$_4$, and C$_2$H$_4$S. This is especially true for M82, an evolved starburst
with main processes driven by UV photons, where all six calculated bright bands were observed.

\subsection{Sulfur-bearing species in galactic molecular gases: Orion KL and Sgr B2}
\label{molmilky}

Orion KL and Sgr B2 are well known galactic molecular clouds where star-forming is pronounced \citep{Goicoechea2008}.
They both were investigated by {\it ISO} in the far-IR region where they are in the group of the brightest sources. 

Orion KL was investigated in the (486-492) GHz and (541-577) GHz intervals by the {\it Odin} satellite, and 280 spectral lines were observed from 38 molecules  \citep{Persson2007}.  Among them 64  lines were unidentified.
Several sulfur species were detected: SO$_2$, SO, H$_2$CS, H$_2$S, OCS, CS, HCS$^+$, as well as formaldehyde. 
Tentative assignments were done for SiS, SH$^-$, and SO$^+$.
Sulfur-bearing species in Orion KL were also studied with the {\it IRAM} 30 m telescope in the (80-115.5) GHz, (130-178) GHz, and (197-281) GHz intervals \citep{Tercero2010,Esplugues2013}. More than 14400 spectral features were detected and 10040 were identified and attributed to 43 molecules, among them to sulfur species: OCS, HCS$^+$, H$_2$CS, CS, CCS, C$_3$S, SO, and SO$_2$. These results \citep{Persson2007,Tercero2010,Esplugues2013} show that rich sulfur chemistry takes place in Orion KL. 
We find that two bands calculated for  S$_5$CH$_2$
(52.65 and 57.72 $\mu$m in our Table~\ref{tab3}) are  close to unidentified bands at 52.57 and 58.02 $\mu$m observed by {\it ISO} in the Orion KL (see their Table 10) \citep{Lerate2006}. However, Lerate and coworkers observed only the  far-IR region (44--188) $\mu$m, whereas the majority of the bands we calculate for pentathian (as well as for other sulfur molecules studied in this work) are in the near and mid-IR. 

It was proposed that similar sulfur chemistry exists in the nucleus of the starburst galaxy NGC 253 and Sgr B2 in the galactic center \citep{Martin2008}. Sgr B2 was studied with {\it ISO} in the near/mid-IR \citep{Lutz1996} and in 
the far-IR region \citep{Goicoechea2004,Polehampton2007}, but the attention was not devoted to the sulfur-bearing species. However, some unidentified 
features were observed \citep{Polehampton2007}. The {\it Mopra} 3-mm spectral line survey confirms that several sulfur-bearing molecules exist in Sgr B2
\citep{Armijos2014}. These are: CCS, OCS, CS, SiS, HCS$^+$, SO$_2$, and HNCS. Several unidentified lines were also observed.
Sulfur-bearing species H$_2$S, SO, SO$_2$, H$_2$CS, OCS, NS, SH$^+$, and HCS$^+$ were observed in Sgr B2 by the {\it Herschel} telescope molecular line survey in the (157--625) $\mu$m range \citep{Neill2014}.

\section{Conclusions}
\label{concl}

Astrochemistry of sulfur is still unknown. 
We calculated IR spectra of several sulfur-bearing molecules and proposed that they take the role in the balance of the sulfur  in dense clouds and comets. We studied ring molecules (CH$_2$S)$_3$, (CH$_2$S)$_4$, 
pentathian (S$_5$CH$_2$), hexathiepan (S$_6$CH$_2$), thiirane (C$_2$H$_4$S), and in addition two other sulfur-bearing molecules
trisulfane (HSSSH) and thioacetone ((CH$_3$)$_2$CS).
We found that calculated bands of (CH$_2$S)$_3$, (CH$_2$S)$_4$, and C$_2$H$_4$S
agree with several unidentified features observed by {\it ISO} 
in starburst  galaxies (M82, NGC 253, the 30 Doradus star-forming region in the Large Magellanic Cloud),
and galaxies with active galactic nuclei (NGC 1068, Circinus).
Studies of molecular compositions of several of these extragalactic objects 
on {\it ALMA} are still in progress. 
Many star-forming and other objects exist, and could act as the place with rich sulfur chemistry. Galactic objects where complex sulfur-bearing
species could easily form are Orion KL and Sgr B2.
We hope that a high resolution and sensitivity of {\it ALMA}, and in the future of the {\it James Webb Space Telescope}
in the infrared spectral region, will enable detection of larger sulfur-bearing  molecules.

\section*{Acknowledgments}
This work  was done using 
the CRO-NGI e-grid and the computational cluster ``Isabella''
at the University of Zagreb Computing Centre SRCE.
G. Bilalbegovi\' c acknowledges the support of the University of Zagreb research fund, grant ``Physics of Stars and Cosmic Dust''.
We are grateful to Guillermo Manuel Mu{\~ n}oz Caro for the useful correspondence, as well as for sending his PhD thesis and the article
\citep{Jimenez2014} prior publication.
This research has made use of NASA's Astrophysics Data System Bibliographic Services. We thank the referee for constructive comments.

\bibliographystyle{mn2e} 
\bibliography{irhcs}
\label{lastpage}

\end{document}